\documentclass[onecolumn,manuscript,showpacs,amsmath,amssymb]{revtex4}
\usepackage{txfonts}
\usepackage{pifont} 
\usepackage{amssymb}
\usepackage{dcolumn}
\usepackage{amsmath}
\usepackage[dvips]{epsfig}

\makeatletter
\def\btt#1{\texttt{\@backslashchar#1}}%
\DeclareRobustCommand\bblash{\btt{\@backslashchar}}%
\makeatother

\begin{document}

\title{Unveiling the Phase Diagram and Nonlinear Optical Responses of a Twisted  Kitaev Chain}
\author{Ya-Min Quan$^{1}$\footnotemark[2],  Shi-Qing Jia$^{2}$, Xiang-Long Yu$^{3}$, Hai-Qing Lin$^{4}$ and Liang-Jian Zou$^{1}$\footnotemark[1]}
 
\footnotetext[1]{ Electronic mail:zou@theory.issp.ac.cn} 
\footnotetext[2]{ Electronic mail:ymquan@theory.issp.ac.cn}

\affiliation{\it $^1$ Key Laboratory of Materials Physics, Institute of Solid State Physics, HFIPS, Chinese Academy of Sciences, Hefei 230031, China\\
         	\it $^2$ Department of Applied Physics, North China University of Science and Technology, Tangshan 063210, China\\
            \it $^3$ School of Science, Sun Yat-sen University, Shenzhen 518107, China\\ 
            \it $^4$ Institute for Advanced Study in Physics and School of Physics, Zhejiang University, Hangzhou, 310058, China  
          } 
\date{today}  

\begin{abstract}

	Detecting Kitaev interactions in real materials remains challenge,
	as conventional experimental techniques often have difficulty distinguishing
	fractionalized excitations from other normal contributions.
	Terahertz two-dimensional coherent spectroscopy (2DCS) offers a novel approach
	for probing many-body phenomena, such as exotic excitations in quantum magnets.
Motivated by recent experiments on CoNb$_2$O$_6$ and the development of the terahertz spectroscopy
 in Kitaev quantum spin liquid, we proposed a twisted Kitaev model for CoNb$_2$O$_6$ and determined
  the precise twist angle according to experimental specific-heat phase diagram. 
  With this calibrated model, we found that 
  non-rephasing diagonal and rephasing anti-diagonal signals appear in the 2DCS  nonlinear  response.
   The $x$ and $y$ components of the spin superexchange interactions split the rephasing 
   signals into a grid of discrete peaks. 
   We further demonstrate that the diagonal 
   and the discrete rephasing signals primarily originate from two-spinon and 
   four-spinon excitation processes based on numerical projection method.  
	These findings indicate that even weak Kitaev interactions in quantum materials
	can be effectively detected via two-dimensional coherent spectroscopy .


\end{abstract}

\maketitle

\section{INTRODUCTION}

Quantum spin liquid (QSL) is a spin disordered phase of quantum matter,
in which the wave function of spins is strongly entangled order and the 
low-energy excitations are dominated by fractionalized particles
\cite{Zhou-RevModPhys-89-025003-2017, 
Savary-RepProgPhys-80-016502-2017,
Balents-Nature-464-199-2010}.  
Kitaev model provides a precious opportunity   to study QSL 
since it is exactly solvable and the ground state shows exotic QSL phase 
\cite{Kitaev-AnnPhys-2-321-2006}. 
In recent years, experimental research on the QSL 
focused on 4d and 5d transition-metal oxides\cite{Takagi-NatRevPhys-1-264-2019}, 
primarily through investigating 
the dynamics of low-energy excitations
\cite{Takagi-NatRevPhys-1-264-2019,
	Wen-npjQuantumMater-4-12-2019, 
	Banerjee-NatMater-15-733-2016, 
	Banerjee-Science-356-1055-2017,
	Banerjee-NpjQuantumMater-3-8-2018,
	Do-NatPhys-13-1079-2017}
 and thermal transport properties 
\cite{Kasahara-PRL-120-217205-2018,
	Tanaka-NatPhys-18-429-2022,
	Czajka-NatPhys-17-915-2021, 
	Yokoi-Science-373-568-2021} 
in candidate QSL materials. 
The broad continuum spectrum observed in the dynamical probe is proposed
as the signature of fractional excitations in the Kitaev QSL.
But the ambiguity of other origin for continuum spectrum can not be excluded 
absolutely\cite{Takagi-NatRevPhys-1-264-2019}.
On the other hand, in the real candidate materials, the Heisenberg interaction and 
other non-diagonal interaction may coexist with the Kitaev interactions, 
leading to magnetic orders at low temperature in some candidate Kitaev  materials 
\cite{Zhou-RevModPhys-89-025003-2017,
	Takagi-NatRevPhys-1-264-2019,
	Banerjee-NatMater-15-733-2016, 
	Eichstaedt-PRB-100-075110-2019}.
 Besides, it is difficult to detect the effects of the Kitaev interaction in 
 these materials by the conventional experimental methods, since
 there is no apparent character of fractional excitation in the spectra signature. 
 Therefore, how to detect the Kitaev interaction in the canditate 
 Kitaev materials remains a great challenge.


 Two-dimensional coherent spectroscopy (2DCS) has been widely used to
study electronic excitation and 
dynamics\cite{Mukamel-1995,Mukamel-AnnuRevPhysChem-51-691-2000,
Jonas-RevPhysChem-54-425-2003}.
In 2DCS, the system is excited by three
laser pulses and the subsequently coherent light emission is measured 
\cite{Phuc-PRB-104-115105-2021,Wan-PRX-11-031035-2021,
Armitage-PRL-122-257401-2019}.
These signals are displayed as a function of the frequencies that 
obtained by performing a Fourier transformation
of the time intervals. 
The diagonal/off-diagonal signals in the 2DCS 
reveal the coupling of the optical excitation and 
emission frequencies\cite{Phuc-PRB-104-115105-2021}.
Since 2DCS could reflect the nonlinear responses due to
 multi-time correlations, it has been widely used to probe the
  nonlinear optical properties of molecular and 
  semiconductor systems\cite{Cundiff-AccChemRes-42-1423-2009,Lu-PRL-118-207204-2017}. 
Recently, this technology is applied to
study strongly correlated many-body systems.   
Since the energy of THz photon falls in a proper energy range of
many-body quasi-partical excitations, and 
the domain of the response in 2DCS is mainly determined by the character of 
quasi-particle excitation, it has  been suggested to
 detect the crisp signatures of spinon and fractional 
 excitations \cite{Armitage-PRL-122-257401-2019,Wan-PRX-11-031035-2021,
Kim-PRL-124-117205-2020,Qiang-PRL-113-126505-2024,Nandkishore-PRR-3-013254-2021}.
The itinerant Majorana fermions and flux excitations contribute 
to the sharp diagonal and anti-diagonal signals in
2DCS\cite{Kim-PRL-124-117205-2020,Armitage-PRL-122-257401-2019}.


  On the other hand,  the quasi-one-dimensional   
  columbite CoNb$_{2}$O$_{6}$ has attracted considerable attention in recent years
  \cite{Cabrera-PRB-90-014418-2014,Kobayashi-PRB-94-134427-2016,
  	Ringler-PRB-105-224421-2022,  
  	Morris-PRL-112-137403-2014,Morris-NatPhys-17-832-2021,
  	Kinross-PRX-4-031008-2014,Liang-NatCommun-6-7611-2015, 
  	Kobayashi-PRB-90-060412-2014}.
  	In this material, the Co$^{2+}$ ions form a zigzag chain along the $c$ axis.
  	The intrachain interaction of  Co$^{2+}$ is ferromagnetic  along the easy axis, while 
  	the  interchain coupling is very weak, leading to a complex magnetic phase 
  	diagram\cite{Thota-PRB-103-064415-2021}.
  	Experimentally, by  applying a magnetic field  along the $b$ axis,  a 
  	quantum critical point (QCP)is observed, marking a transition from a spin-ordered phase to a paramagnetic
  	state  at a critical field 
  	$B_{c}\approx 5.3T$
  	\cite{Coldea-Science-327-177-2010,Cabrera-PRB-90-014418-2014,Sachdev-1999,Thota-PRB-103-064415-2021}. 
  	Near the quantum critical point, the spectrum of bound states
  	 exhibits a structure corresponding to the E$_{8}$ root in the presence of a perturbing longitudinal field\cite{Coldea-Science-327-177-2010}. 
  	 So it was proposed as an ideal model system of the transverse-field Ising chain
  	\cite{Liang-NatCommun-6-7611-2015,Kinross-PRX-4-031008-2014}.
Meanwhile, some obvious notable deviations from the transverse-field
  Ising model(TFIM) behaviors displays away from the critical
   point\cite{Fava-PNAS-117-25219-2020,Morris-NatPhys-17-832-2021}.
To account for these discrepancies,the magnon excitation spectrum observed via inelastic 
neutron scattering is typically described using an XXZ model 
incorporating NN and next-nearest-neighbor (NNN) interactions within a single cobalt spin chain
\cite{Cabrera-PRB-90-014418-2014,Woodland-PRB-108-184417-2023,Gallegos-PRB-109-014424-2024}. 
 Recently, a staggered off-diagonal exchange term has been proposed to explain the soliton 
spectrum\cite{Fava-PNAS-117-25219-2020,Woodland-PRB-108-184417-2023,Gallegos-PRB-109-014424-2024}. 
In the quantum paramagnetic phase, the off-diagonal exchange interactions 
has also been used to explain the magnon decay and spectrum 
renormalization \cite{Gallegos-PRB-109-014424-2024}. 
Additionally, Morris et al. proposed a twisted Kitaev model(TKM) with bond dependent 
Ising interactions for CoNb$_{2}$O$_{6}$ to  interpret the deviations of absorption 
THz spectroscopy data from ideal TFIM\cite{Morris-NatPhys-17-832-2021}.
This simplified model includes only the nearest-neighbor (NN) Ising-type coupling, neglecting spin exchange along 
the $x$ and $y$ directions. 
   However, more comprehensive spin model that incorporates additional interactions
   is needed to reproduce the curvature of the 
   magnon dispersion of CoNb$_{2}$O$_{6}$\cite{Dagotto-PRB-107-104414-2023}.    
   All the results demonstrate that the off-diagonal interactions primarily originate from the structural 
  twist angle and significantly influence spinon (kinks, or domain wall) domain wall dynamics. 
  Therefore, incorporating this twist angle 
  into the NN XXZ model enables describing the spinons  dynamics in CoNb$_{2}$O$_{6}$.  
  Moreover, the ground state of CoNb$_{2}$O$_{6}$ is ferromagnet and 
 it is difficult to obtain clear evidence of bond-dependent interactions from 
 thermal transport measurements\cite{Liang-NatCommun-6-7611-2015} or
  neutron scattering experiment\cite{Kobayashi-PRB-94-134427-2016}.
This raises the questions of whether significant Kitaev interactions are present or not
in this material, and how such bond-dependent interactions affect quasiparticle excitations.
  2DCS is ideal for the study of exictations in quantum materials
  and it will play a key role in addressing this problem.

In this work, we explore 2DCS signatures 
of a TKM developed for describing the quantum material CoNb$_{2}$O$_{6}$. 
We first construct a modified TKM model by introducing a twist angle into an anisotropic 
nearest-neighbor(NN) Heisenberg framework to characterize the magnetic interactions in CoNb$_{2}$O$_{6}$. 
The twist angle is determined by comparing theoretical predictions with experimental specific heat measurements. 
To identify the signals induced by the Kitaev interaction,  we numerically investigate the 2DCS of the modified TKM  
using the exact diagonalization method. 
We analyze the diagonal and anti-diagonal spectral features in 2DCS that arise from finite twist angles 
 and the effect of NN  XY exchange terms on these spectra,
 which leads to the discreteness of the rephasing spectrum (spin echo signals). 
Furthermore, the contributions of various multi-spinon excitations to the spectral signals are identified  
via projection methods.
 It is indicated that the two-spinon excitation is the most important process to 
the 2DCS. 
Our findings demonstrate that 2DCS serves as an effective tool for probing Kitaev interactions in quantum materials.
  The rest of this paper is organized as follows.
  In Sec. \uppercase\expandafter{\romannumeral2} 
  we introduce a realistic TKM 
  for CoNb$_{2}$O$_{6}$ and 
  the 2DCS formalism, as well as the munerical methods used to 
  compute the 2DCS based on Lanczos algorithm.
  In Sec. \uppercase\expandafter{\romannumeral3} 
  we present the numerical results and discuss the Kitaev interaction
  features in 2DCS, the conclusion is drawn 
  in Sec. \uppercase\expandafter{\romannumeral4}.


\section{MODEL HAMILTONIAN AND NUMERICAL METHOD}

 \begin{figure}[htbp]
	\centering 
    \includegraphics[angle=0, width=0.8 \columnwidth]{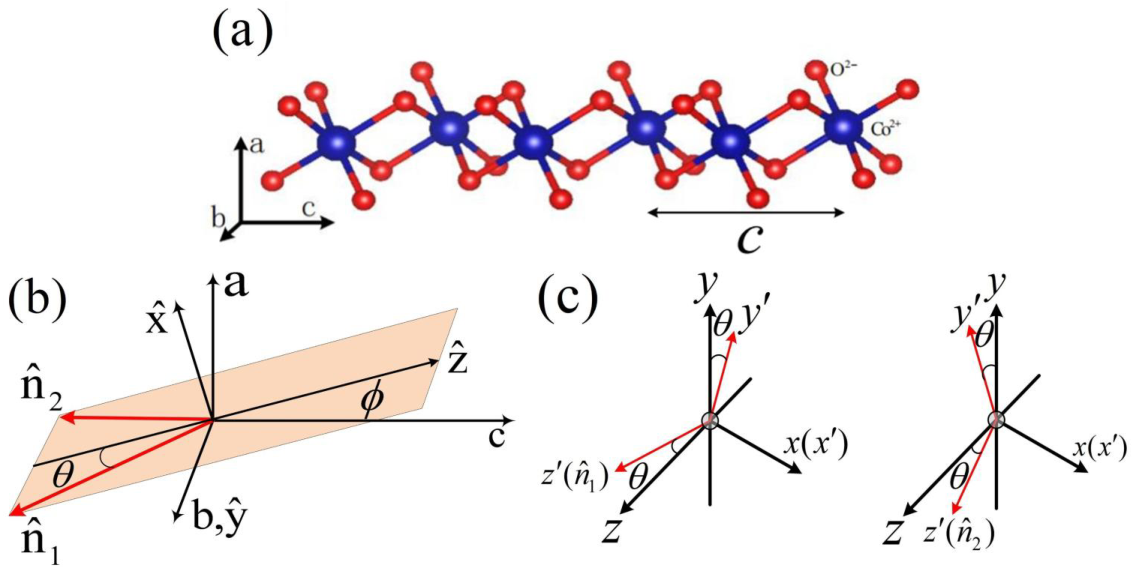}
	\caption{(Color online) 
		(a) The twisted Kitaev chain of Co$^{2+}$ ions and distorted O$^{2-}$ octahedra\cite{Morris-NatPhys-17-832-2021}.
	The crystalgraphic a,b, and c directions are indicated. 
	(b) $\hat{n}_{1}$ and $\hat{n}_{2}$ are the two alternating Ising direcitons.
	The $\hat{y}$ spin direction is identified with the b axis.
	$\theta$ is the twist angle that $\hat{n}_{1}$ and $\hat{n}_{2}$ made with 
	the $\hat{z}$ axis. (c)	The laboratory reference frame $[x,y,z]$ and the local reference frame 
	$[x^{\prime},y^{\prime},z^{\prime}]$ with z axis along the spin orientation.
	} 	   
	\label{Fig1:structure} 
\end{figure}

CoNb$_{2}$O$_{6}$ crystallizes in an orthorhombic structure with space group \textit{Pbcn}. 
Magnetic Co$^{2+}$ ions form one-dimensional (1D) chains in the ac plane, 
extending along the crystallographic c-axis. These chains exhibit a staggered shift along the b-axis, 
as depicted in Fig. \ref{Fig1:structure}. 
Within the basal ab plane, the Co$^{2+}$ spins form a weakly coupled, distorted triangular lattice. 
 The combined effects of Hund's coupling and spin-orbit coupling (SOC) stabilize 
 a $J_{\text{eff}}=1/2$ state for the Co$^{2+}$ ions\cite{Heid-JMMM-151-123-1995}. 
 Neutron diffraction reveals a ground-state magnetic structure below $1.97$ K 
 characterized by ferromagnetic ordering  moments within each chain and antiferromagnetic ordering between chains. 
 Experimentally, the easy magnetization axes of the Co$^{2+}$ ions are tilted from the c axis by approximately ±30°, 
 with the sign of this tilt angle alternating between adjacent chains.  
 To facilitate analysis using 2DCS, 
 our work employs a simplified distorted Kitaev model considering only NN exchange interactions between cobalt ions. 
 As shown in Fig. \ref{Fig1:structure}(b), we define a laboratory coordinate system $[x,y,z]$
 by rotating the easy magnetization axis by $\gamma=30$\textdegree around the b-axis, aligning the z-axis  
 with the Ising easy direction. Due to the twist angle, we establish two local coordinate systems 
 $[x^{\prime},y^{\prime},z^{\prime}]$ based on the distinct coupling directions $\hat{n}_{1}$
 and $\hat{n}_{2}$ to project local interactions. 
 These local frames are obtained by rotating the laboratory system 
 by the twist angle $\theta$ around the x-axis respectively.  
  We consider the anisotropic Heisenberg Hamiltonian in these local coordinates:
 
 \begin{eqnarray}   
 	\label{eq:Heisenberg}
 	H_{\text{local}}&=&-J_{x^{\prime}}\sum_{i} s_{i}^{x^{\prime}} s_{i+1}^{x^{\prime}}
 	-J_{y^{\prime}}\sum_{i} s_{i}^{y^{\prime}} s_{i+1}^{y^{\prime}}
 	-J_{z^{\prime}}\sum_{i} s_{i}^{z^{\prime}} s_{i+1}^{z^{\prime}},   		 
 \end{eqnarray}  
 where $J_{x^{\prime}}$, $J_{y^{\prime}}$ and $J_{z^{\prime}}$ are the NN magnetic 
 exchange constants in the local frames. 
 By transforming the model into the lab frame, we obtain the Hamiltonian for 
 the TKM for NN exchange interactions as follows, see Appendix \ref{app:TKmodel}:   
 \begin{eqnarray}   
	\label{eq:Hamiltonian1}
	H_{\text{TKM}}&=&-J_{x^{\prime}}\sum_{i} s_{i}^{x} s_{i+1}^{x} \nonumber\\ 
	& &-J_{y^{\prime}}\sum_{i}\left[\sin^{2}(\theta)s_{i}^{z} s_{i+1}^{z}
	+\cos^{2}(\theta)s_{i}^{y} s_{i+1}^{y} 
	-\dfrac{\sin(2\theta)}{2}(-1)^{i}\left(s_{i}^{z} s_{i+1}^{y}
	+s_{i}^{y} s_{i+1}^{z} \right) \right]  \nonumber\\ 	
	& &-J_{z^{\prime}}\sum_{i}\left[\cos^{2}(\theta)s_{i}^{z} s_{i+1}^{z}
	+\sin^{2}(\theta)s_{i}^{y} s_{i+1}^{y} 
	+\dfrac{\sin(2\theta)}{2}(-1)^{i}\left(s_{i}^{y} s_{i+1}^{z}
	+s_{i}^{z} s_{i+1}^{y} \right) \right]
	-h\sum_{i}s_{i}^{y},	 	 	 
\end{eqnarray} 
here  $\theta$ is the twist angle between the two NN $\hat{n}$ directions, where 
$\hat{n}$ is the bond-dependent Ising interaction direction as shown 
in Fig.\ref{Fig1:structure}(c), $h$ is the transverse magnetic field. 
  At $\theta=0$\textdegree, the Ising nature of the Hamiltonian 
 is dominant. 
 The Kitaev nature becomes dominant at $\theta=45$\textdegree, which
 is not intuitive since the Kitaev interaction is  the off-diagonal 
 terms\cite{Dagotto-PRB-107-104414-2023}.  
 Comparison with the established exchange constant matrices  yields the local coordinate exchange constants for 
 CoNb$_{2}$O$_{6}$: 
 $J_{x^{\prime}}=0.57$ meV, $J_{y^{\prime}}=0.51$ meV, and $J_{z^{\prime}}=2.64$ meV \cite{Gallegos-PRB-109-014424-2024}.  
 The unknown twist angle $\theta$ will be determined by fitting the experimental
 results of specific heat\cite{Liang-NatCommun-6-7611-2015}.


   Next, we present the formalism of 2DCS for the twisted kitaev model.   
   Based on the experimental procedure of 2DCS, 
   we consider probing the Kitaev interaction in a spin system 
   using two pulsed magnetic fields separated by a time interval $\tau_{1}$. 
   The higher-order nonlinear response arising from the coherence induced by 
   these pulses is measured after a further delay $\tau_{2}$ following the second pulse. 
   Setting the time of the first pulse at $t=0$, the second pulse occurs 
   at $t=\tau_{1}$. The $\alpha$-component pulsed magnetic field signal can be 
   written as \cite{Kim-PRL-124-117205-2020,Armitage-PRL-122-257401-2019}:
\begin{eqnarray}
	\label{eq:nonlinear_signal}
	B^{\alpha}(t)=B^{\alpha}_{0}\delta(t)+B^{\alpha}_{1}\delta(t-\tau_{1}),
\end{eqnarray}
   The nonlinear magnetization response $M^{\alpha}_{NL}$ in the $\alpha$-direction,
    measured at $t=\tau_{1}+\tau_{2}$, is defined as the difference between 
    the total magnetization induced by both pulses and the sum of the magnetizations induced by each pulse individually:
$ M^{\alpha}_{NL}(\tau_{1}+\tau_{2})=M^{\alpha}_{B^{\alpha}_{0}B^{\alpha}_{1}}(\tau_{1}+\tau_{2})
-M^{\alpha}_{B^{\alpha}_{0}}(\tau_{1}+\tau_{2})-M^{\alpha}_{B^{\alpha}_{1}}(\tau_{1}+\tau_{2})$.
   Here, $M^{\alpha}_{NL}$ represents the nonlinear magnetization resulting from the coherent 
   interaction of the two magnetic pulses, $M^{\alpha}_{B^{\alpha}_{0}B^{\alpha}_{1}}$ is 
   the magnetization induced by both pulses, and 
   $M^{\alpha}_{B^{\alpha}_{0}}$, $M^{\alpha}_{B^{\alpha}_{1}}$ are 
   the  responses to the individual pulses $B^{\alpha}_{0}$ and $B^{\alpha}_{1}$, respectively. 
   Expanding this expression in terms of linear and nonlinear 
   susceptibilities yields \cite{Kim-PRL-124-117205-2020,Armitage-PRL-122-257401-2019}:
\begin{eqnarray} 
	\label{eq:nonlinear_response}
	M^{\alpha}_{NL}(\tau_{2}+\tau_{1})/2L&=&\chi^{(2)}_{\alpha\alpha\alpha}(\tau_{2},\tau_{1})B^{\alpha}_{\tau_{1}}B^{\alpha}_{0} 
	+\chi^{(3)}_{\alpha\alpha\alpha\alpha}(\tau_{2},\tau_{1},0)B^{\alpha}_{\tau_{1}}B^{\alpha}_{0}B^{\alpha}_{0} \nonumber\\
	& &+\chi^{(3)}_{\alpha\alpha\alpha\alpha}(\tau_{2},0,\tau_{1})B^{\alpha}_{\tau_{1}}B^{\alpha}_{\tau_{1}}B^{\alpha}_{0}
	+O(B^{4}), 
\end{eqnarray}  
   where $L$ is the number of sites. The $n$-th order susceptibility is given by \cite{Mukamel-1995,Kim-PRL-124-117205-2020}:
\begin{eqnarray}  
	\label{eq:nonlinear_susceptibility1}
	\chi^{(n)}_{\alpha\cdots\alpha}(\tau_{n},\cdots,\tau_{1})&=&\dfrac{i^{n}}{L} 
	\varTheta(\tau_{1})\varTheta(\tau_{2})\cdots \varTheta(\tau_{n})  \nonumber\\
	& &\langle \left[ \left[ \cdots \left[ \hat{M}^{\alpha}(\tau_{n}+\cdots+\tau_{1}),
	\hat{M}^{\alpha}(\tau_{n-1}+\cdots+\tau_{1}) \right], \cdots\right],\hat{M}^{\alpha}(0) \right]  \rangle.
\end{eqnarray}   
  Due to the slip symmetry of the system, the correlation function component 
   in the second-order susceptibility vanishes under slip symmetry 
   transformation \cite{Sim-PRB-108-134423-2023}. Consequently, 
   this work focuses primarily on the third-order susceptibility. 
   Expanding the commutators in Eq. (\ref{eq:nonlinear_susceptibility1}) 
   yields the following expressions for the third-order susceptibility:
   \begin{eqnarray}
   	\label{eq:tdns_third1}
   	\chi^{(3)}_{\alpha\alpha\alpha\alpha}\left( \tau_{2},\tau_{1},0 \right)&=& -\dfrac{2}{L}
   	\varTheta(\tau_{1})\varTheta(\tau_{2})
   	\sum_{l=1}^{4} \mathrm{Im} \left[ R^{(l)}_{\alpha\alpha\alpha\alpha}\left( \tau_{2},\tau_{1},0 \right)\right] ,
   \end{eqnarray}
   \begin{eqnarray}
   	\label{eq:tdns_third2}
   	\chi^{(3)}_{\alpha\alpha\alpha\alpha}\left( \tau_{2},0,\tau_{1} \right)&=& -\dfrac{2}{L}
   	\varTheta(\tau_{1})\varTheta(\tau_{2})
   	\sum_{l=1}^{4} \mathrm{Im} \left[ R^{(l)}_{\alpha\alpha\alpha\alpha}\left( \tau_{2},0,\tau_{1} \right)\right] .
   \end{eqnarray}
   The higher-order correlation functions $R^{(l)}_{\alpha\alpha\alpha\alpha}\left( t_{3},t_{2},t_{1} \right)$ 
   in Eqs. (\ref{eq:tdns_third1}) and (\ref{eq:tdns_third2}) are defined as:
   \begin{eqnarray}
	\label{eq:tdns_third3}
	R^{(1)}_{\alpha\alpha\alpha\alpha}\left( t_{3},t_{2},t_{1} \right)=\langle 
	\hat{M}^{\alpha}(t_{3}+t_{2}+t_{1}) \hat{M}^{\alpha}(t_{2}+t_{1})
	\hat{M}^{\alpha}(t_{1})\hat{M}^{\alpha}(0)
	\rangle,
\end{eqnarray}    
\begin{eqnarray}
	\label{eq:tdns_third4}
	R^{(2)}_{\alpha\alpha\alpha\alpha}\left( t_{3},t_{2},t_{1} \right)=-\langle 
	\hat{M}^{\alpha}(t_{2}+t_{1}) \hat{M}^{\alpha}(t_{3}+t_{2}+t_{1}) 
	\hat{M}^{\alpha}(t_{1})\hat{M}^{\alpha}(0)
	\rangle.
\end{eqnarray} 
\begin{eqnarray}
	\label{eq:tdns_third5}
	R^{(3)}_{\alpha\alpha\alpha\alpha}\left( t_{3},t_{2},t_{1} \right)=-\langle 
	\hat{M}^{\alpha}(t_{1})	 \hat{M}^{\alpha}(t_{3}+t_{2}+t_{1}) 
	\hat{M}^{\alpha}(t_{2}+t_{1}) \hat{M}^{\alpha}(0)
	\rangle,
\end{eqnarray} 
\begin{eqnarray}
	\label{eq:tdns_third6}
	R^{(4)}_{\alpha\alpha\alpha\alpha}\left( t_{3},t_{2},t_{1} \right)=\langle 
	\hat{M}^{\alpha}(t_{1}) \hat{M}^{\alpha}(t_{2}+t_{1})	 
	\hat{M}^{\alpha}(t_{3}+t_{2}+t_{1}) \hat{M}^{\alpha}(0)
	\rangle,
\end{eqnarray} 
   In this work, we compute the correlation functions 
   $R^{(l)}_{\alpha\alpha\alpha\alpha}\left( t_{3},t_{2},t_{1} \right)$ 
   using the Lanczos method as seen in Appendix \ref{app:Lanczos}.
    The standard Lanczos method generates an $M_{L}$-dimensional Krylov subspace, 
    within which we solve for $M_{L}$ eigenstates. The correlation functions are 
     then evaluated within this subspace.  
     For example, $R^{(1)}_{\alpha\alpha\alpha\alpha}\left( \tau_{2},0,\tau_{1} \right)$ becomes:
 \begin{eqnarray} 
 	\label{eq:tdns_third11}
 	R^{(1)}_{\alpha\alpha\alpha\alpha}\left(\tau_{2},0,\tau_{1} \right)=
 	\dfrac{1}{Z}\sum_{n}e^{-\beta \varepsilon_{n}}\sum_{pqv=0}^{M_{L}} 
 	\langle \Psi_{n} | \hat{M}^{\alpha}|\Psi_{p}\rangle  	
 	\langle \Psi_{p} | \hat{M}^{\alpha} |\Psi_{q}\rangle \nonumber\\    
 	\times 	\langle \Psi_{q} | \hat{M}^{\alpha} |\Psi_{v}\rangle 
 	\langle \Psi_{v} | \hat{M}^{\alpha} |\Psi_{n}\rangle
 	e^{i(\varepsilon_{n}-\varepsilon_{v})\tau_{1}}e^{i(\varepsilon_{n}-\varepsilon_{p})\tau_{2}},  	 	 
 \end{eqnarray} 
   where the states $|\Psi_{n}\rangle$ to $|\Psi_{v}\rangle$ are 
   eigenvectors computed via the orthogonal Lanczos method and $Z$ is the partition function.  
   See Appendix \ref{app:Lanczos} for the details of the simulation.

\section{ NUMERICAL RESULTS AND DISCUSSION}

    \subsection{Determining the Twist Angle in TKM via Specific Heat Data}

   \begin{figure}[htbp]
   	\centering
   	\includegraphics[angle=0, width=0.45 \columnwidth]{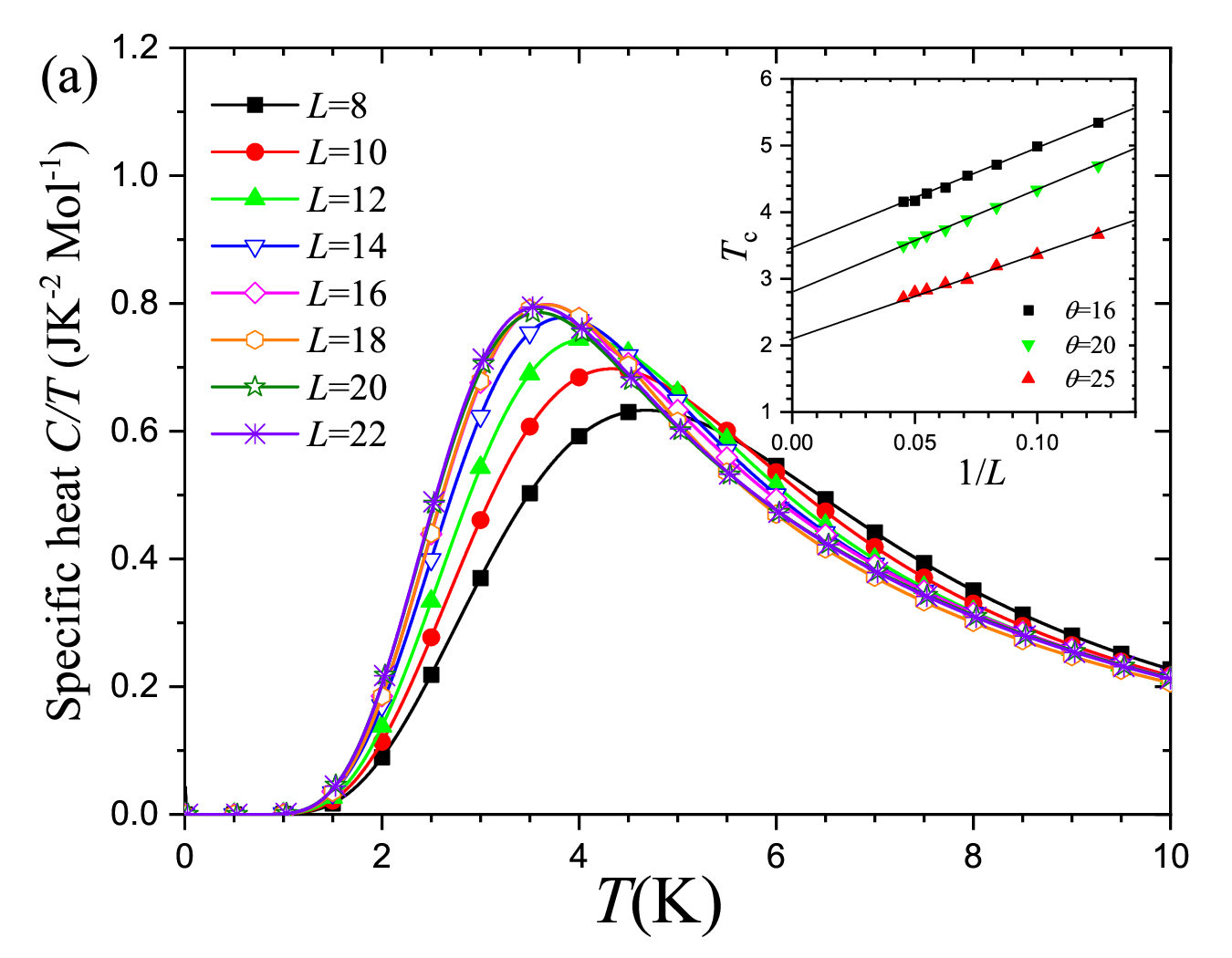} 
   	\includegraphics[angle=0, width=0.45 \columnwidth]{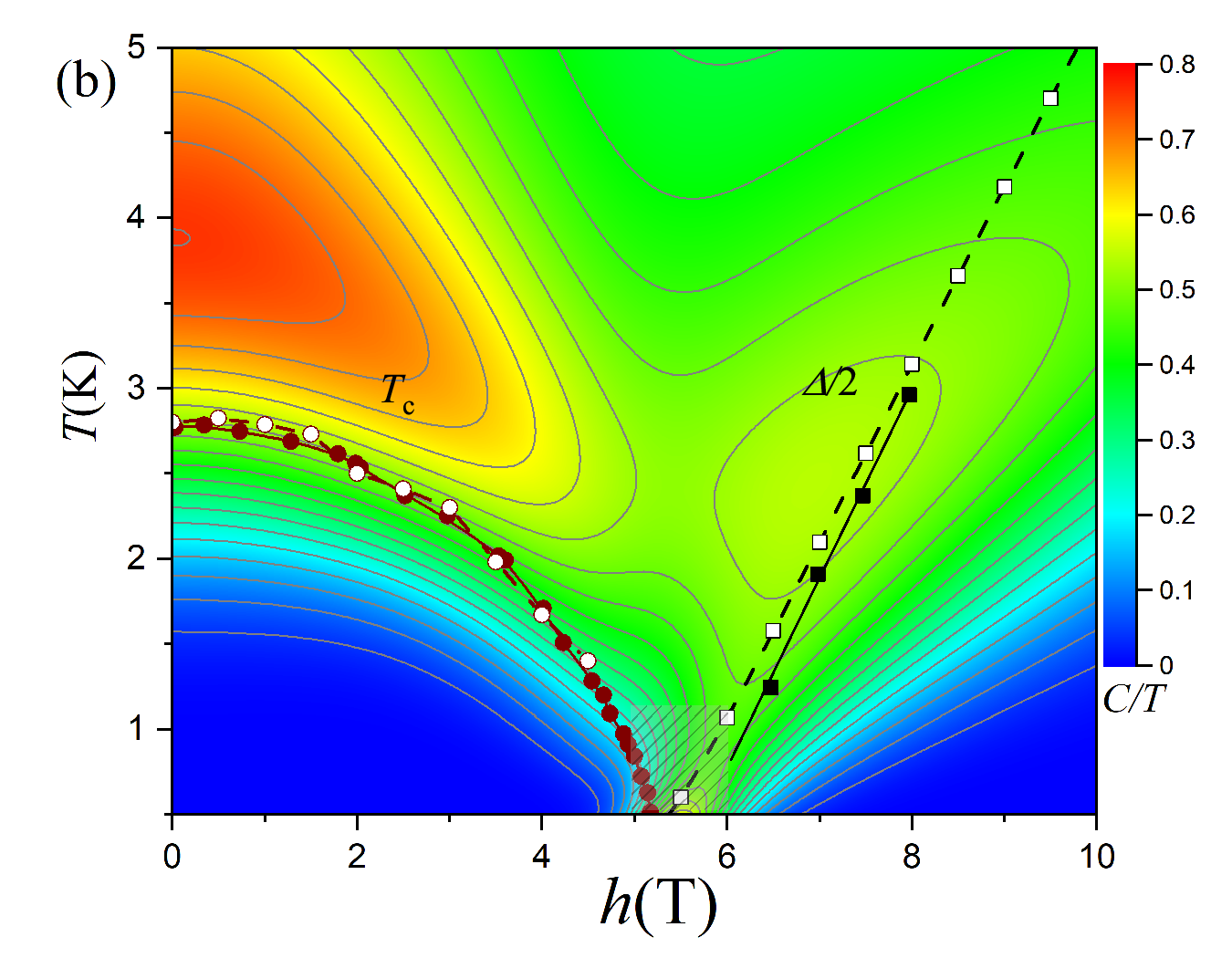}
   	\caption {(Color online)   		
   	(a) The specific heat  of the TKM for 
   	different chain length at $\theta$=20\textdegree.
   	    Insite is the finite-size scaling result.   	    
   	 (b) The phase diagrams of the tiwsted Kitaev model inferred 
   	    from $C/T$ with $\theta$=20\textdegree and chain length $L=18$.
   	    The solid lines with circle and square symbols are the 
   	    transition temperature $T_{c}$ and 
   	    gap $\Delta$\cite{Liang-NatCommun-6-7611-2015}. 
   	    The dash lines with hollow circle and square symbols
   	    are the finite-size extrapolated numerical
   	    results for comparison.} 
   	\label{fig2:capacity}   
   \end{figure}

    In the TKM, the twist angle governs the relative strength
     of the Kitaev couplings, which in turn profoundly affects both the ground state
      and excitation properties of the system. 
       To achieve more reliable results, 
       we further validate the values of the twist angle by incorporating 
       constraints from thermodynamic experiments.
    The specific heat of the TKM as a function of temperature is plotted in Fig. \ref{fig2:capacity}.
    It is shown that the transition temperature is approximately $2.8$ K and the specific heat 
    vanishes below $1$ K at a twist angle of $\theta=20$\textdegree, 
    which is in consistent with experimental results \cite{Liang-NatCommun-6-7611-2015}.
    As illustrated in the inset of Fig. \ref{fig2:capacity}(a), 
    we computed the system-size dependence of the phase transition temperature, 
    identified by the specific heat peak $C/T$, for various twist angles 
    to account for finite-size effects. 
    With $\theta=16$\textdegree, the transition temperature increases to 3.4 K, significantly 
    exceeding the experimental value. These findings support a twist angle of $\theta=20$\textdegree for CoNb$_{2}$O$_{6}$.
    Fig. \ref{fig2:capacity}(b) presents a phase diagram in the 
    temperature–transverse magnetic field plane based on specific heat data. 
    The critical field for the quantum phase transition from ferromagnetic to 
    paramagnetic states is highly sensitive to the $g$-factor.
    Owing to strong spin–orbit coupling in CoNb$_{2}$O$_{6}$, the $g$-factor is substantially larger than 2,
    with reported values for Co$^{2+}$ ions typically ranging from 2.6 to 3.6
    \cite{Fava-PNAS-117-25219-2020,Woodland-PRB-108-184416-2023,
     Woodland-PRB-108-184417-2023,Gallegos-PRB-109-014424-2024,
     Churchill-PRL-133-056703-2024,Thota-PRB-103-064415-2021}. 
     To achieve better agreement with experiment, we employ a g-factor of 3.0 in  
      Fig. \ref{fig2:capacity}(b). 
    The solid line with circles 
    denotes the experimental phase transition temperature $T_{c}$,
    while the line with squares indicates the experimental gap\cite{Liang-NatCommun-6-7611-2015}. 
    The finite-size extrapolated numerical results of the 
    transition temperature and the excitation gap are also plotted for comparison. 
    For $\theta=20$\textdegree, the boundary lines agree with experiment data very well.
    The TKM model accurately describes 
    the transition from magnetically ordered to paramagnetic states, 
    with a quantum critical point (QCP) at about $h_{c}=5$ T\cite{Liang-NatCommun-6-7611-2015}.     
   The elementary excitations of TKM are spinons (kinks or domain wall) with the  transverse magnetic field below 
   the critical field $h_{c}$ and the elementary excitations becomes magnon-like spin flips above $h_{c}$
   \cite{Coldea-Science-327-177-2010,Gallegos-PRB-109-014424-2024,Watanabe-PRB-110-134443-2024}.

  \begin{figure}[htbp]
  	\centering
  	\includegraphics[angle=0, width=0.45 \columnwidth]{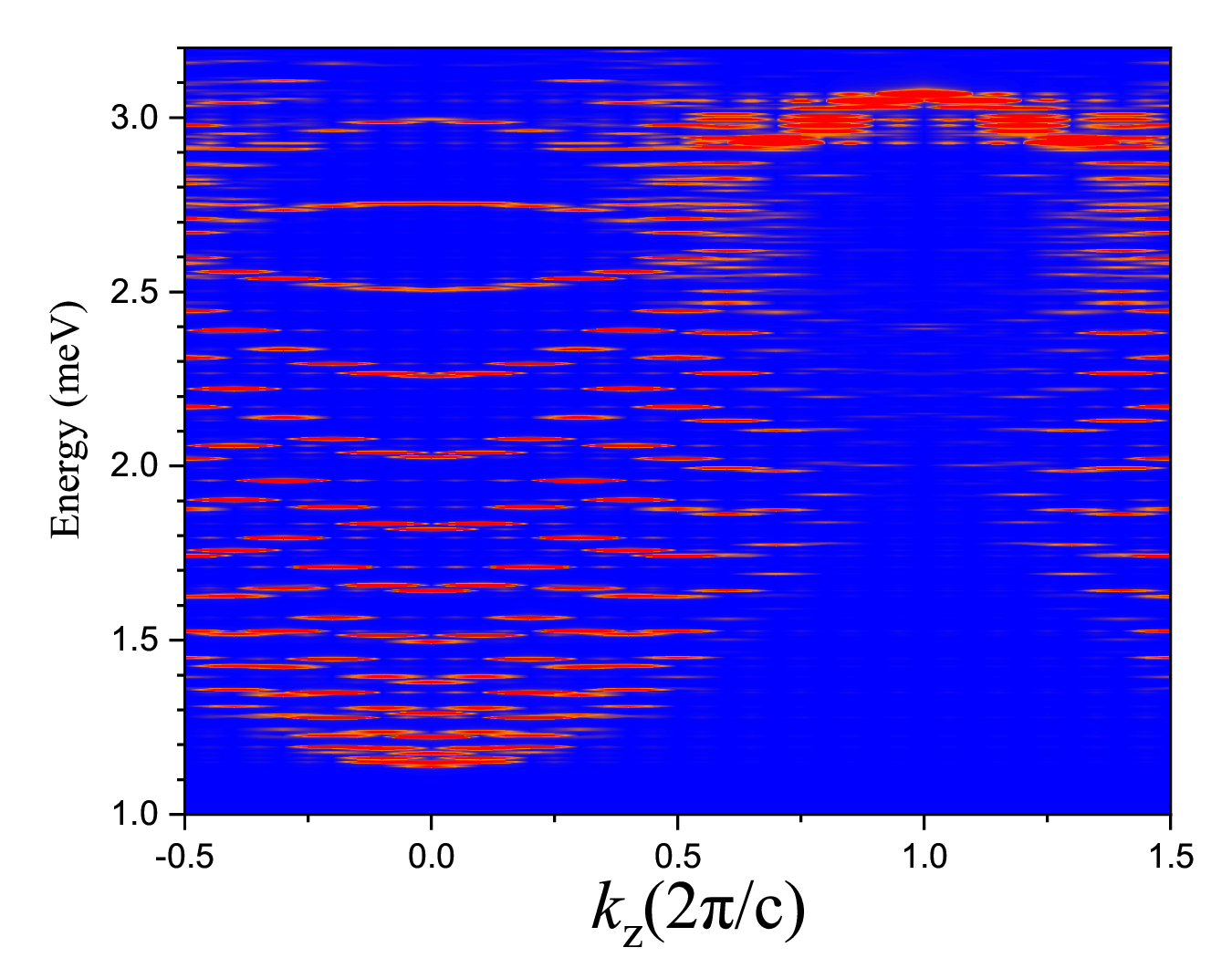} 
  	\caption {(Color online)   
  		$S^{xx}(k,w)$ obtained by the exact diagonalization on a 20-site cluster with periodic
  		boundary conditions at transverse fields $h=0$
  		} 
  	\label{figgpdx2:dispersion}  
  \end{figure}

The dynamic spin structure factor defined as 
 $S^{xx}(q,\omega)=\sum_{v} |\langle \Psi_{v} | S^{x}(q) | \Psi_{0} \rangle|^{2}\delta(\varepsilon_{v}-\varepsilon_{0}-\omega)$
with twist angle $\theta=20$\textdegree is shown in Fig. \ref{figgpdx2:dispersion}.
In Fig. \ref{figgpdx2:dispersion}(a), 
a gapped continuum scattering arising from spinon excitations appears 
near $k=0$ in the energy range from $1.0$ meV to $3.0$V meV. 
The spectrum of the spinon excitations exhibit  a quadratic dispersion shape
due to off-diagonal interactions\cite{Coldea-Science-327-177-2010,Woodland-PRB-108-184416-2023,Woodland-PRB-108-184417-2023}. 
All these results establish that this model describes the spinon excitations in CoNb$_{2}$O$_{6}$      
with only NN interactions.
The above results demonstrate that the dispersions of  spinons
in this material, along with properties like specific heat and 
magnetic field-induced phase transitions, 
can be described by the TKM given in Eq. \eqref{eq:Hamiltonian1}.

       \begin{figure}[htbp]  
   	\includegraphics[angle=0, width=0.45 \columnwidth]{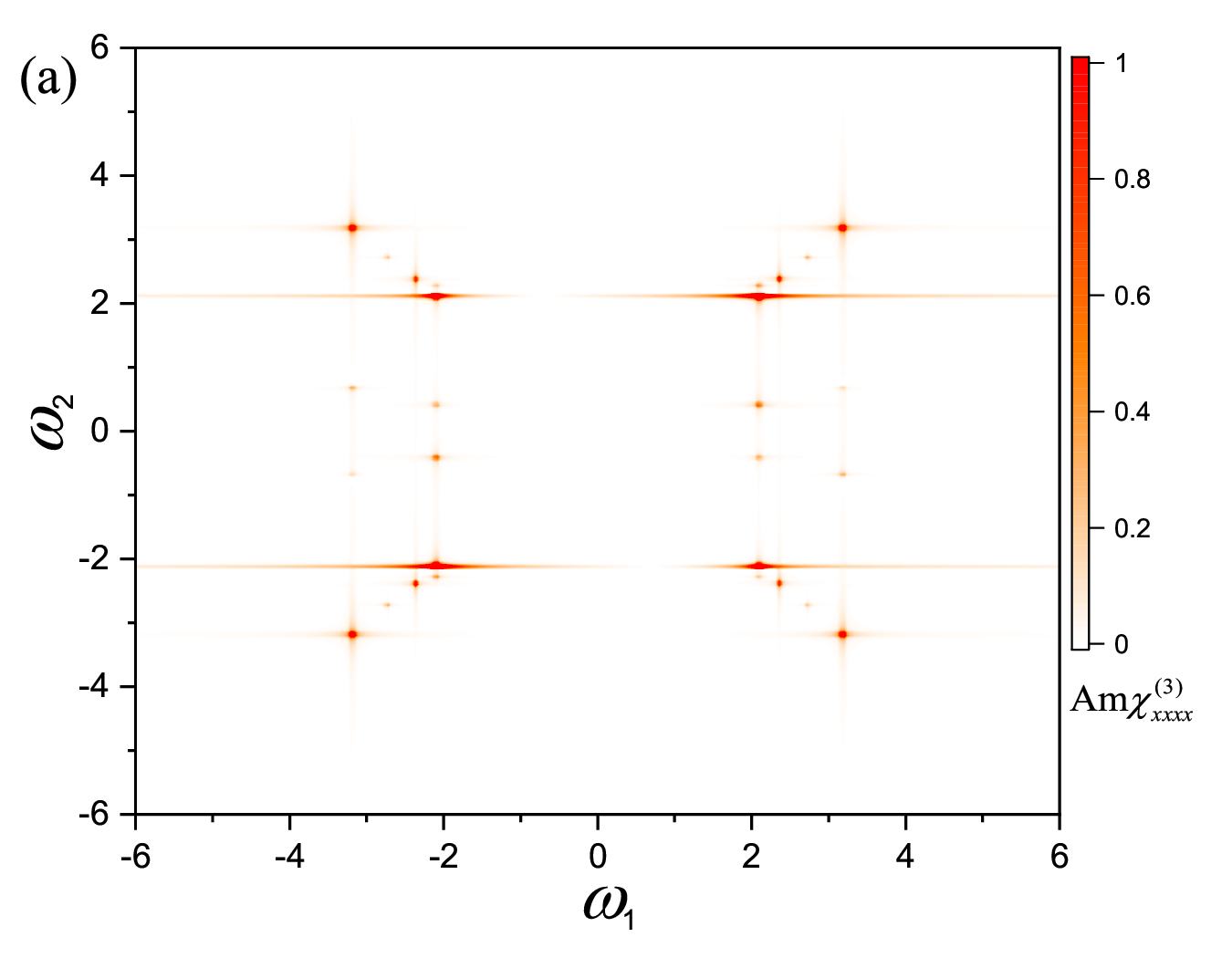} 
   	\includegraphics[angle=0, width=0.45 \columnwidth]{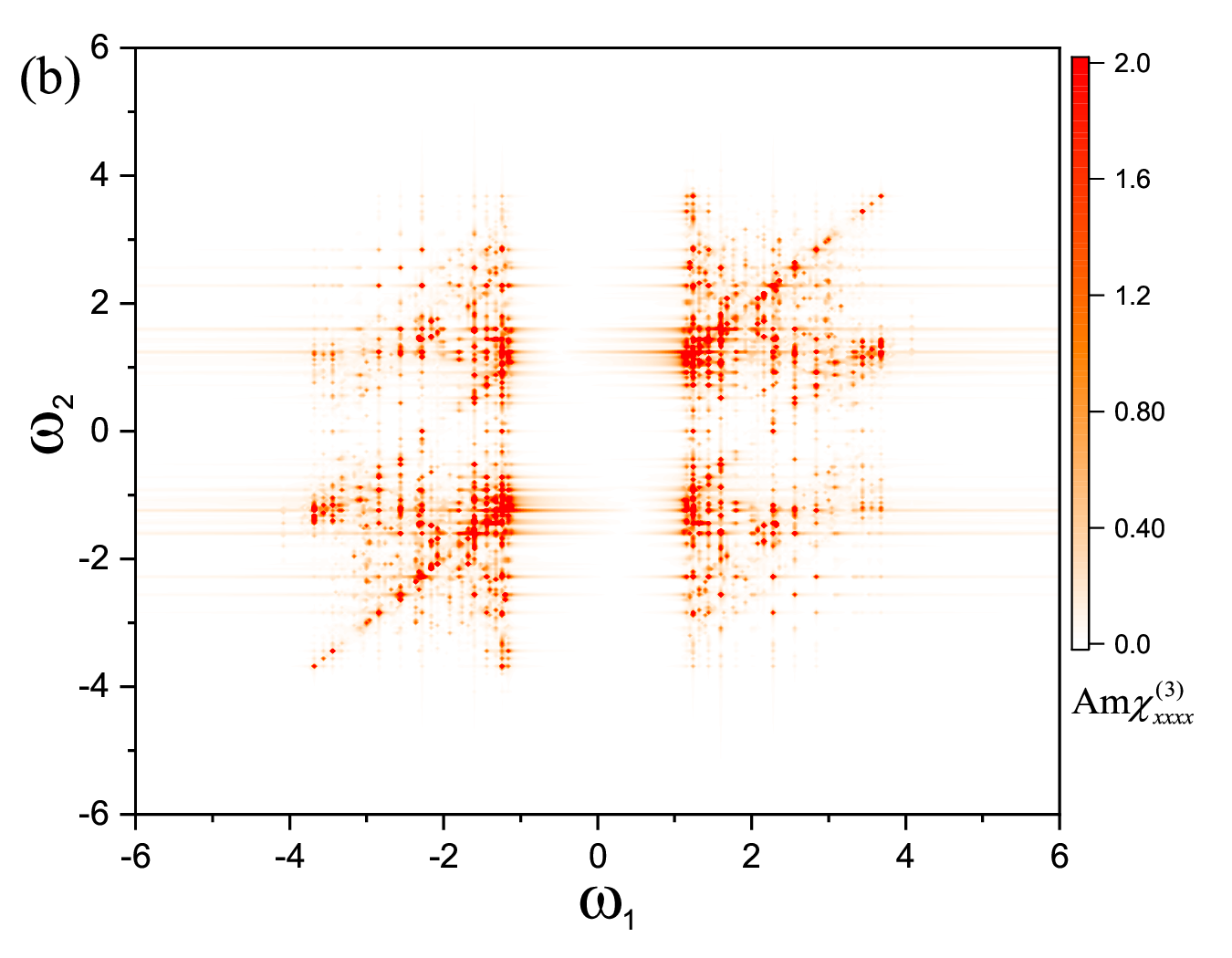} 
   	\caption {(Color online) 
   		Two-dimensional amplitude spectrum of the third-order susceptibilities
   		$\chi^{(3)}_{xxxx}(\omega_{2},0,\omega_{1})$for $\theta$=0\textdegree (a)
   		and $\theta$=20\textdegree (b), respectively with chain length $L=18$. 
   		.}  
   	\label{fig3:tdns3}
   \end{figure}

  \subsection{Two-dimensional coherent spectroscopy of finite twist angle}

 Our next goal is to detect the Kitaev components in CoNb$_{2}$O$_{6}$ through the 2DCS. 
The two-dimensional Fourier spectrums 
of the third-order susceptibilities 
$\chi^{(3)}_{xxxx}(\omega_{2},0,\omega_{1})$ 
are plotted in Fig. \ref{fig3:tdns3}.
For comparision, we analyze the 2DCS at twist angle $\theta=0$\textdegree firstly.
At $\theta=0$\textdegree the ground state of the 
Hamiltonian in Eq. (\ref{eq:Hamiltonian1}) is ferromagnetic and 
the low-energy excitations are dominated by spinon. 
Thus, the spectral features near $\omega_{1} = \omega_{2} = \pm 2$ originate primarily 
from two-spinon single spin-flip excitation processes as shown in Figure \ref{fig3:tdns3}(a)\cite{Nandkishore-PRR-3-013254-2021}. 
 Within the Liouville space, the dominant contribution to these signal peaks 
 arises from the paths that the quantum states $|\Psi_{p}\rangle$ and $|\Psi_{v}\rangle$ 
  contain one spin flip, while the intermediate state $|\Psi_{q}\rangle$ contains either zero or two spin flips.
The spectral signal near $\omega_{2} = 0$  is contributed by bound two-spinon and two spin flips excitations.  
The corresponding Liouville space path is that each of the states $|\Psi_{p}\rangle$ and $|\Psi_{v}\rangle$  
contains one spin flip, while the intermediate state $|\Psi_{q}\rangle$ contains two spin flips\cite{Nandkishore-PRR-3-013254-2021}.
 Unlike the 2D spectrum of Ising model, clustered discrete signals instead of a single one are 
 observed near $(\omega_{1},\omega_{2})=( \pm 2, 0)$,
 which is induced by the finite value of $J_{x}$ and $J_{y}$.

     In contrast, the  shape of the 2DCS changes sharply with $\theta=20$\textdegree for realistic parameters in CoNb$_{2}$O$_{6}$.
     As shown in Fig.\ref{fig3:tdns3}(b), the energy gap decreases to $1.06$ meV. 
      Furthermore, the  finite twist angle
      leads to sharp non-rephasing signals along the diagonal in the first quadrant 
      and it also gives rise to discrete  
      spectral features in the fourth quadrants, which is different from the results given by  Sim and Watanabe  et al. \cite{Watanabe-PRB-110-134443-2024,Sim-PRB-108-134423-2023}.  
In their study,  
sharp rephasing signals were observed along the anti-diagonal 
in the fourth quadrant with only $J_z$ exchange interaction ($J_x = J_y = 0$). 
 The first reason is that  the  twist angle $\theta=20$\textdegree of CoNb$_{2}$O$_{6}$
 is significantly smaller. Secondly, the finite $J_x$ and $J_y$ exchange
interactions in our model cause the rephasing signals to manifest as discrete  structures,
since the spinons become confinment bound states with finite $J_x$ and $J_y$ interactions \cite{Watanabe-PRB-110-134443-2024,Coldea-Science-327-177-2010,Rutkevich-JSatMech-07015-177-2010} .  
   Additionally, the  characteristics of both the non-rephasing 
   signal in the first quadrant and the discrete dot-like rephasing signals in the fourth quadrant show no significant variation with
    increasing system size, demonstrating that these signals are not finite-size 
    effects, as seen Fig.\ref{fig12:tdns_scaling} in Appendix \ref{app:nonsus}.

     \begin{figure}[htbp]
   	\centering
  	\includegraphics[angle=0, width=0.45 \columnwidth]{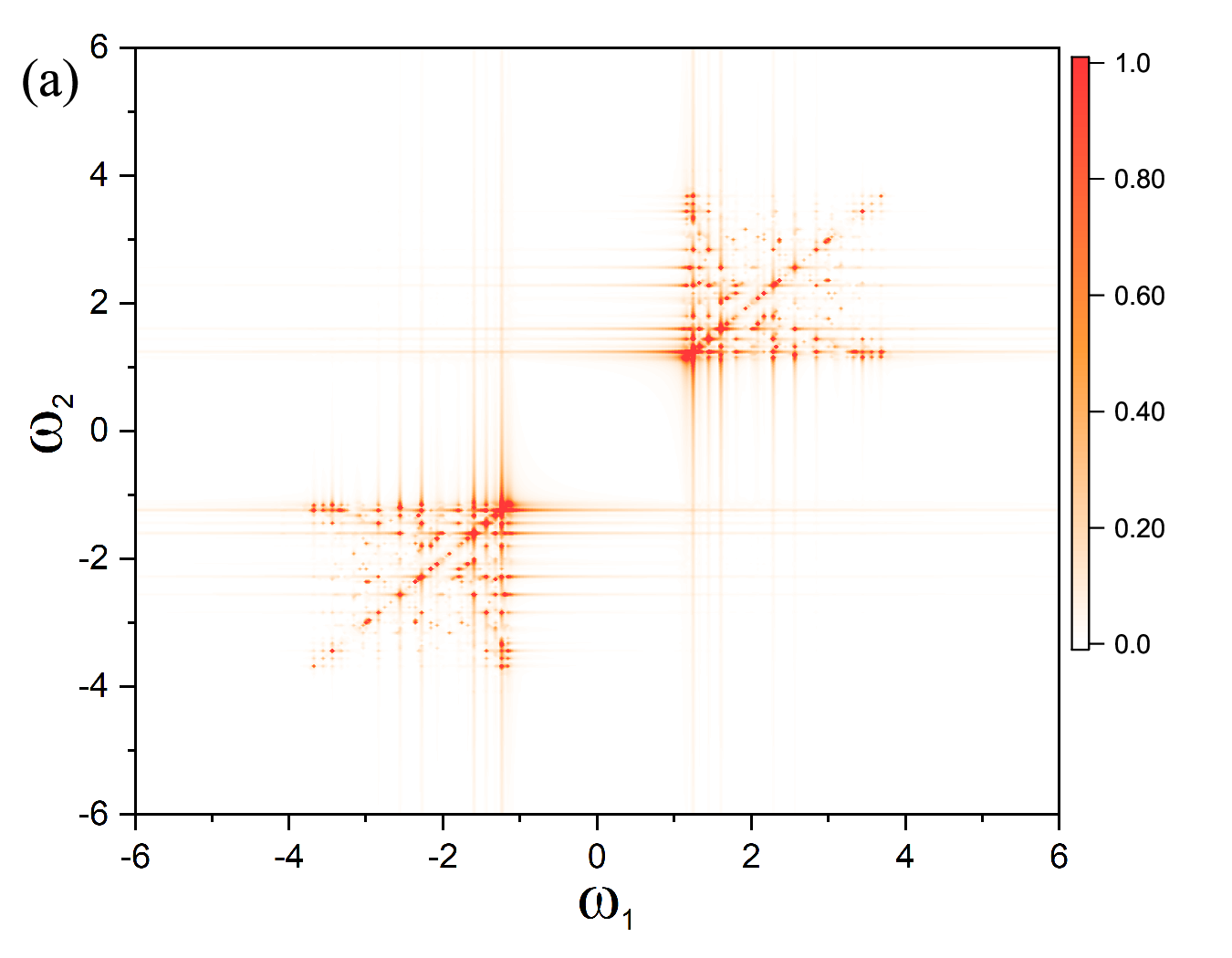} 
  	\includegraphics[angle=0, width=0.45 \columnwidth]{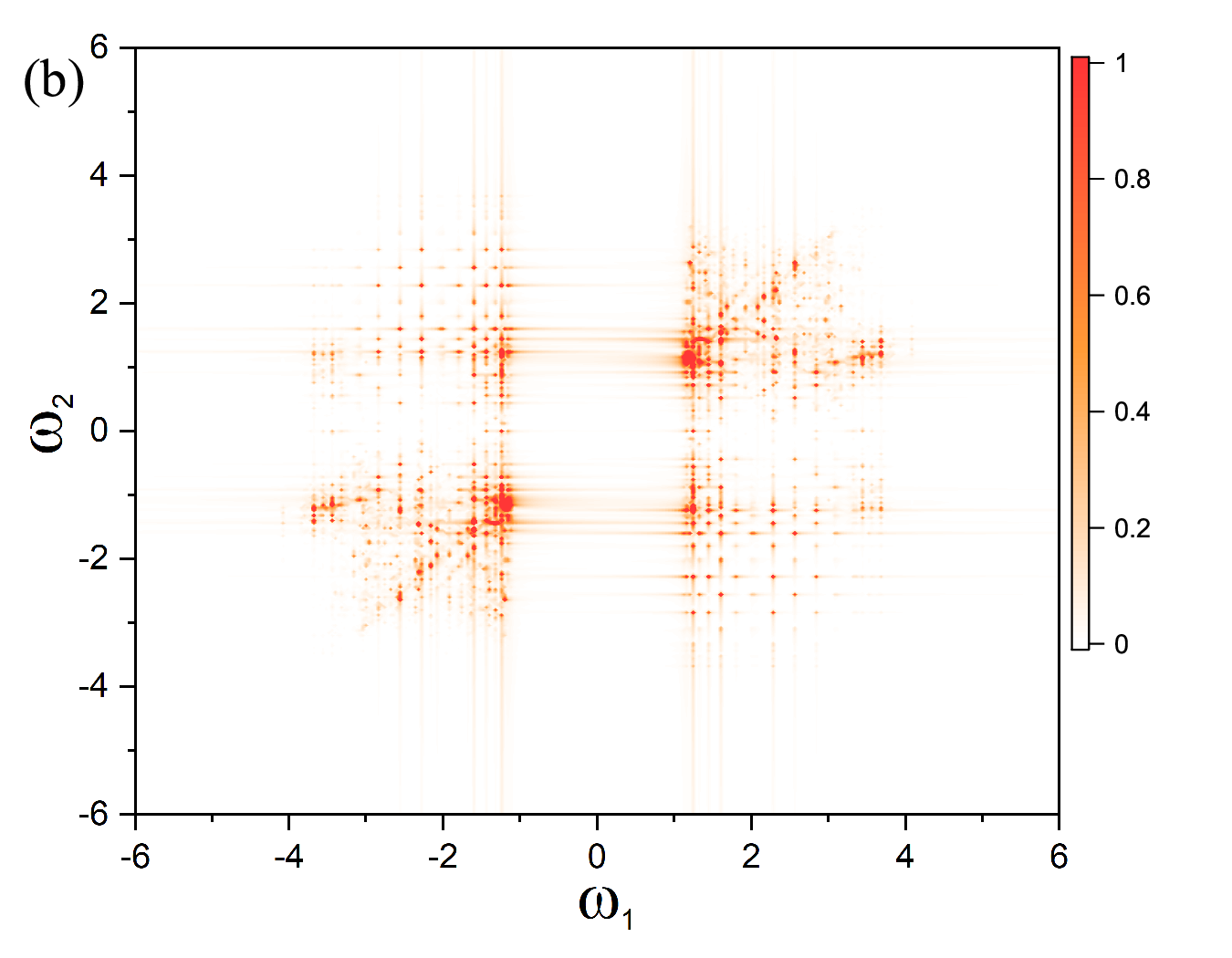}
  	\includegraphics[angle=0, width=0.45 \columnwidth]{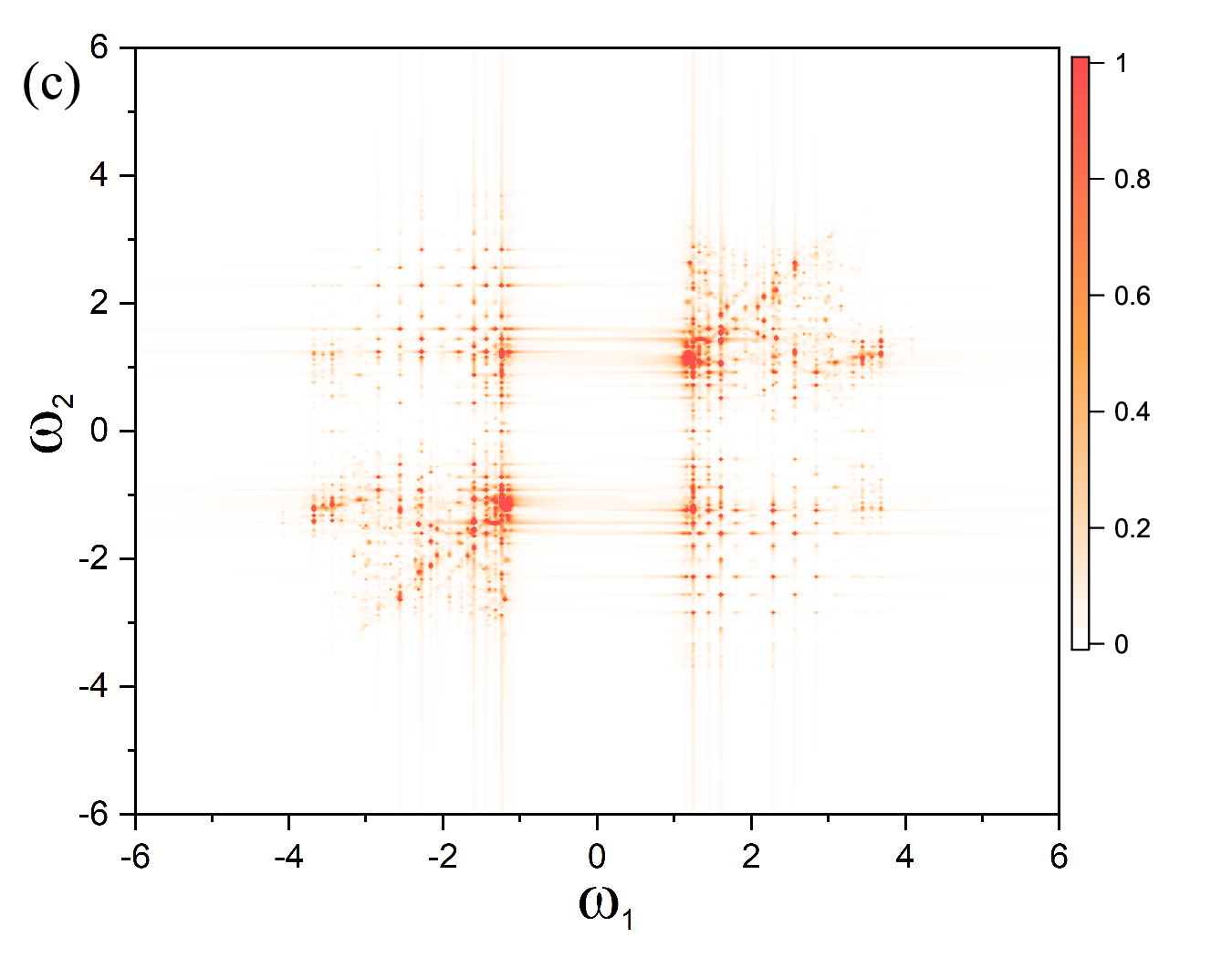} 
  	\includegraphics[angle=0, width=0.45 \columnwidth]{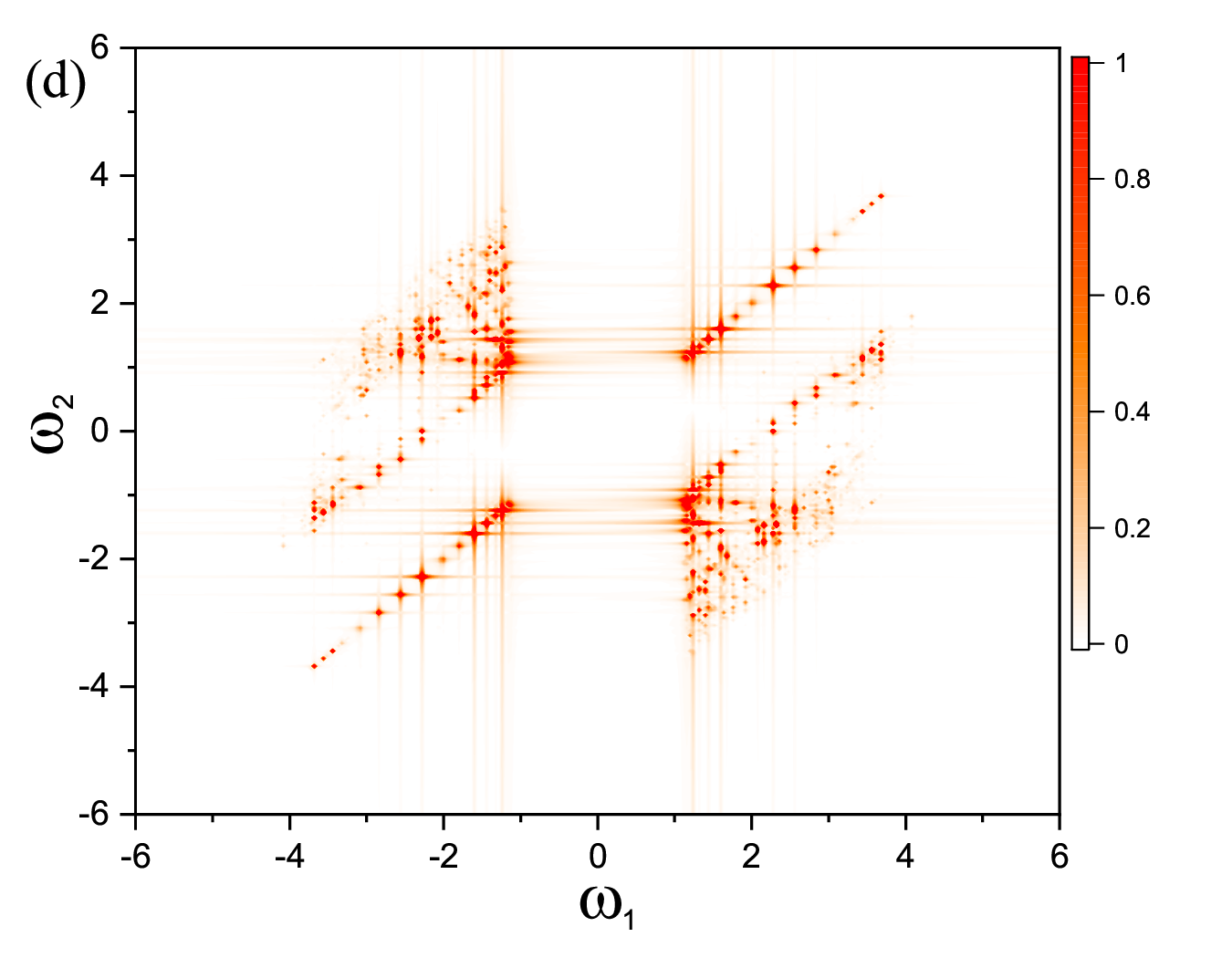}  	
  	\caption {(Color online) 
  		Two-dimensional amplitude spectrum of the four point correlation 
  		functions with $\theta$=20\textdegree and chain length $L=18$,
  		(a)  Am$\mathcal{F}[\text{Im}R_{xxxx}^{(1)}(\tau_{2},0,\tau_{1})]$   
  		(b)  Am$\mathcal{F}[\text{Im}R_{xxxx}^{(2)}(\tau_{2},0,\tau_{1})]$  
  		(c)  Am$\mathcal{F}[\text{Im}R_{xxxx}^{(3)}(\tau_{2},0,\tau_{1})]$  
  		(d)  Am$\mathcal{F}[\text{Im}R_{xxxx}^{(4)}(\tau_{2},0,\tau_{1})]$.
  		Here $\mathcal{F}$ is the Fourier transformation.
	} 		
  	\label{fig4:tdns4} 
   \end{figure}

To identify which processes contribute the rephase and non-rephase sigals in the susceptibilities,
we study the Fourier transformation of $R$ terms in Eq.(\ref{eq:tdns_third2}) respectively.     
 The terms of the third-order susceptibility in
  Eq.(\ref{eq:tdns_third2})  can be described 
 by four distinct double-sided Feynman diagrams 
 (see Fig.\ref{fig11:Feynman_diagram} in Appendix \ref{app:nonsus}) \cite{Watanabe-PRB-110-134443-2024,Mukamel-1995}. 
 To investigate the response processes of spinon excitations, 
  the Fourier-transform spectras of the 
 four-point correlation functions corresponding to these four diagrams are plotted in Fig. \ref{fig4:tdns4}. 
 As shown in Fig. \ref{fig4:tdns4}(a) and (d), $R_{xxxx}^{(1)}(\tau_{2},0,\tau_{1})$
  and  $R_{xxxx}^{(4)}(\tau_{2},0,\tau_{1})$  contribute to non-rephasing signals. 
  In Fig. \ref{fig4:tdns4}(a), the dominant low-energy excitation processes for diagonal non-rephasing 
  signals is that both the excited states $|\Psi_{p}\rangle$ and $|\Psi_{v}\rangle$ 
  contain a pair of spinons, with the intermediate state $|\Psi_{q}\rangle$
  being excited state ($|\Psi_{q}\rangle \neq |\Psi_{n}\rangle $). 
  For $R_{xxxx}^{(4)}(\tau_{2},0,\tau_{1})$ in Fig. \ref{fig4:tdns4}(d),the diagonal non-rephasing signals in the first quadrant 
  mainly attributed by the processes that the intermediate state$|\Psi_{q}\rangle$
  is the ground state, while $|\Psi_{p}\rangle$ and $|\Psi_{v}\rangle$ are both two-kink single spin flip excited states. 
   The noise signals distributed in the fourth quadrant of Fig. \ref{fig4:tdns4}(d) 
  come from the four-spinon excitation  processes with $|\Psi_{q}\rangle \neq |\Psi_{n}\rangle $.
 The channel  $R_{xxxx}^{(2)}(\tau_{2},0,\tau_{1})$ and 
   $R_{xxxx}^{(3)}(\tau_{2},0,\tau_{1})$ contribute equally to the  
  $\chi^{(3),}_{xxxx}(\omega_{2},0,\omega_{1})$ as shown in Fig. \ref{fig4:tdns4}(b) and (c) respectively.
  According to the double-sided Feynman diagrams as seen in  Fig.\ref{fig11:Feynman_diagram} in Appendix \ref{app:nonsus},
The $R_{xxxx}^{(2)}(\tau_{2},0,\tau_{1})$ and 
$R_{xxxx}^{(3)}(\tau_{2},0,\tau_{1})$  are the primary contributors to the discrete rephase signals.  
 The dominant excitation processes responsible for the rephasing signals involve 
 a pathway with the state $|\Psi_{q}\rangle=|\Psi_{n}\rangle$.

  \begin{figure}[htbp]
  	\centering
  	\includegraphics[angle=0, width=0.45 \columnwidth]{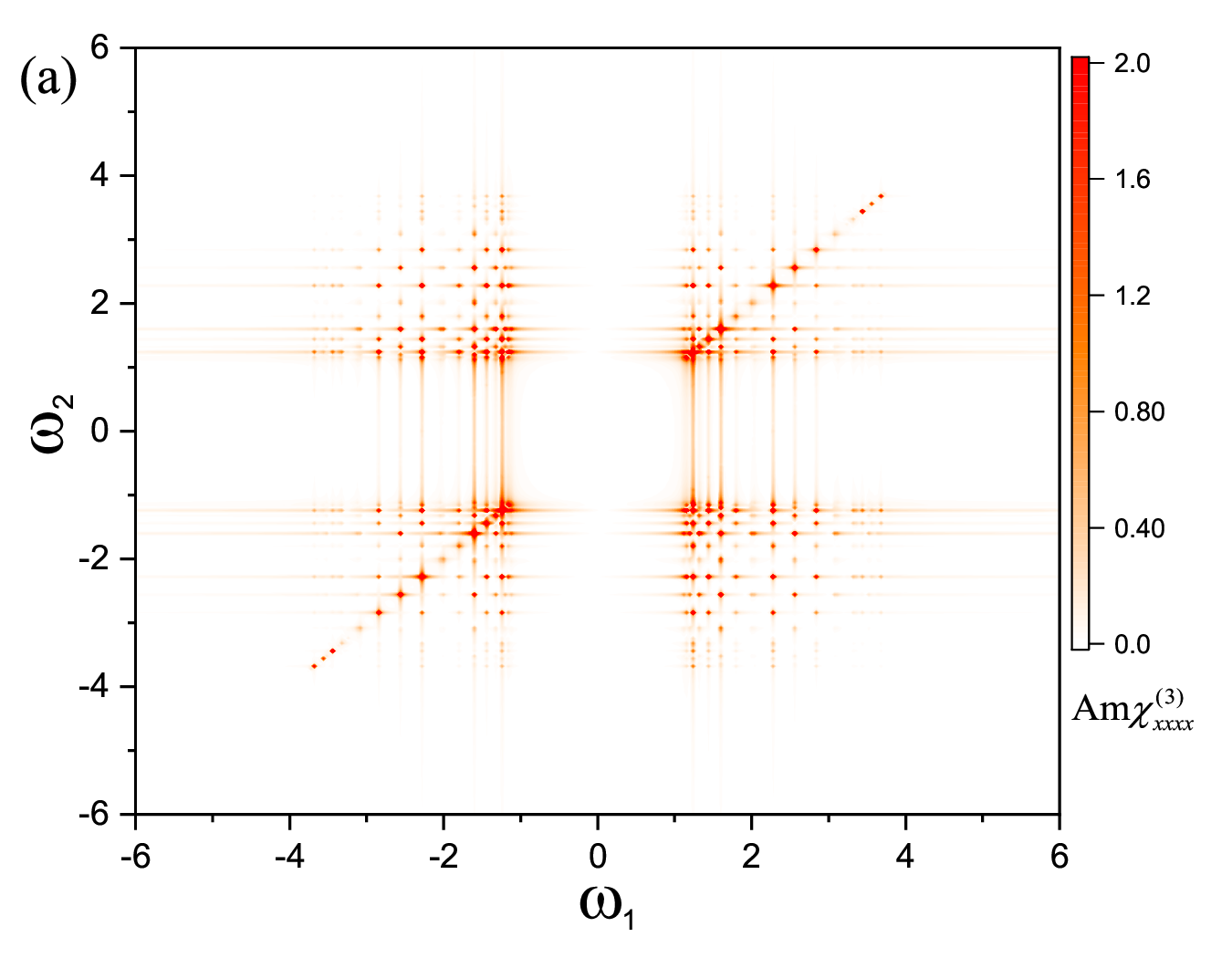} 
  	\includegraphics[angle=0, width=0.45 \columnwidth]{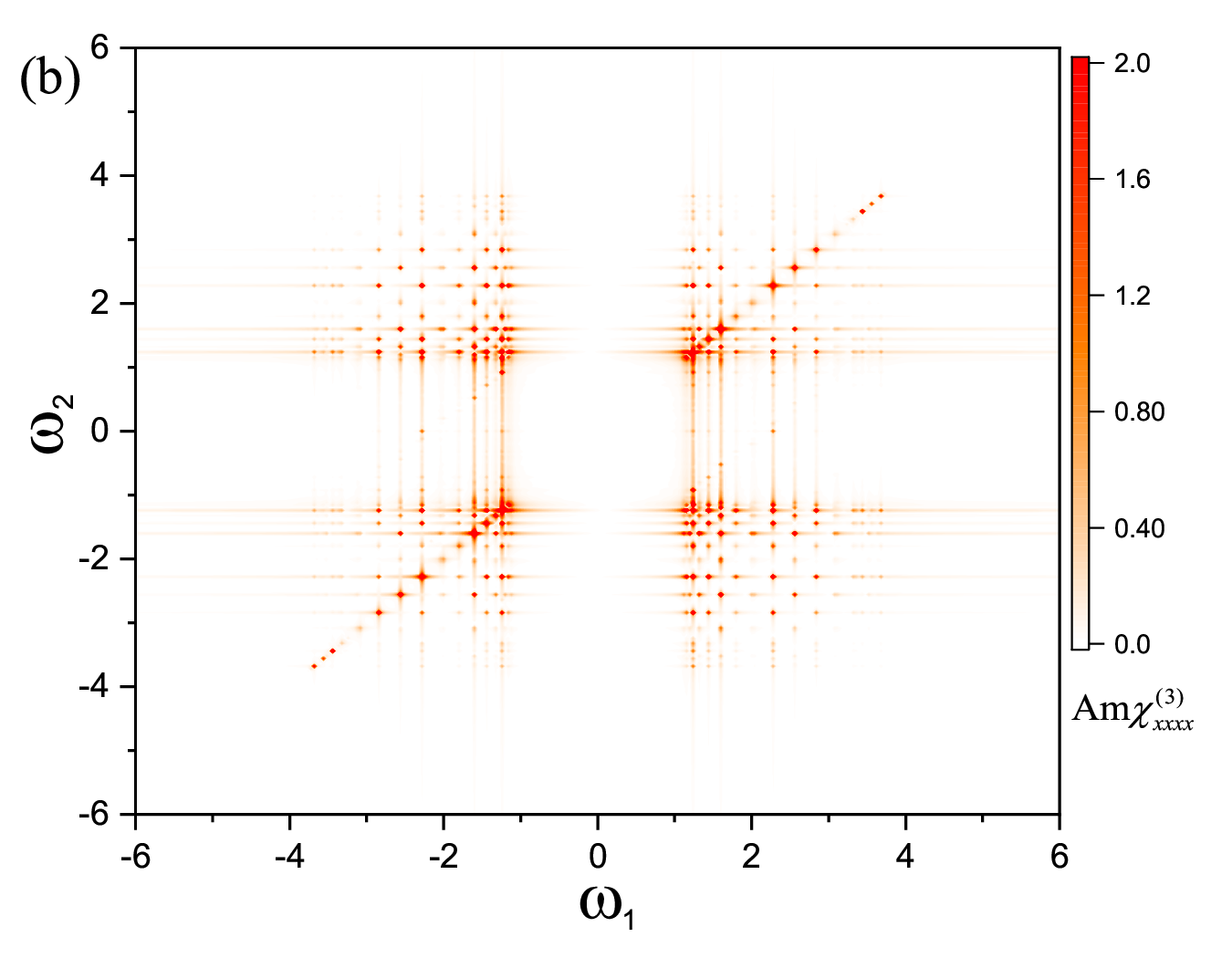}
  	\includegraphics[angle=0, width=0.45 \columnwidth]{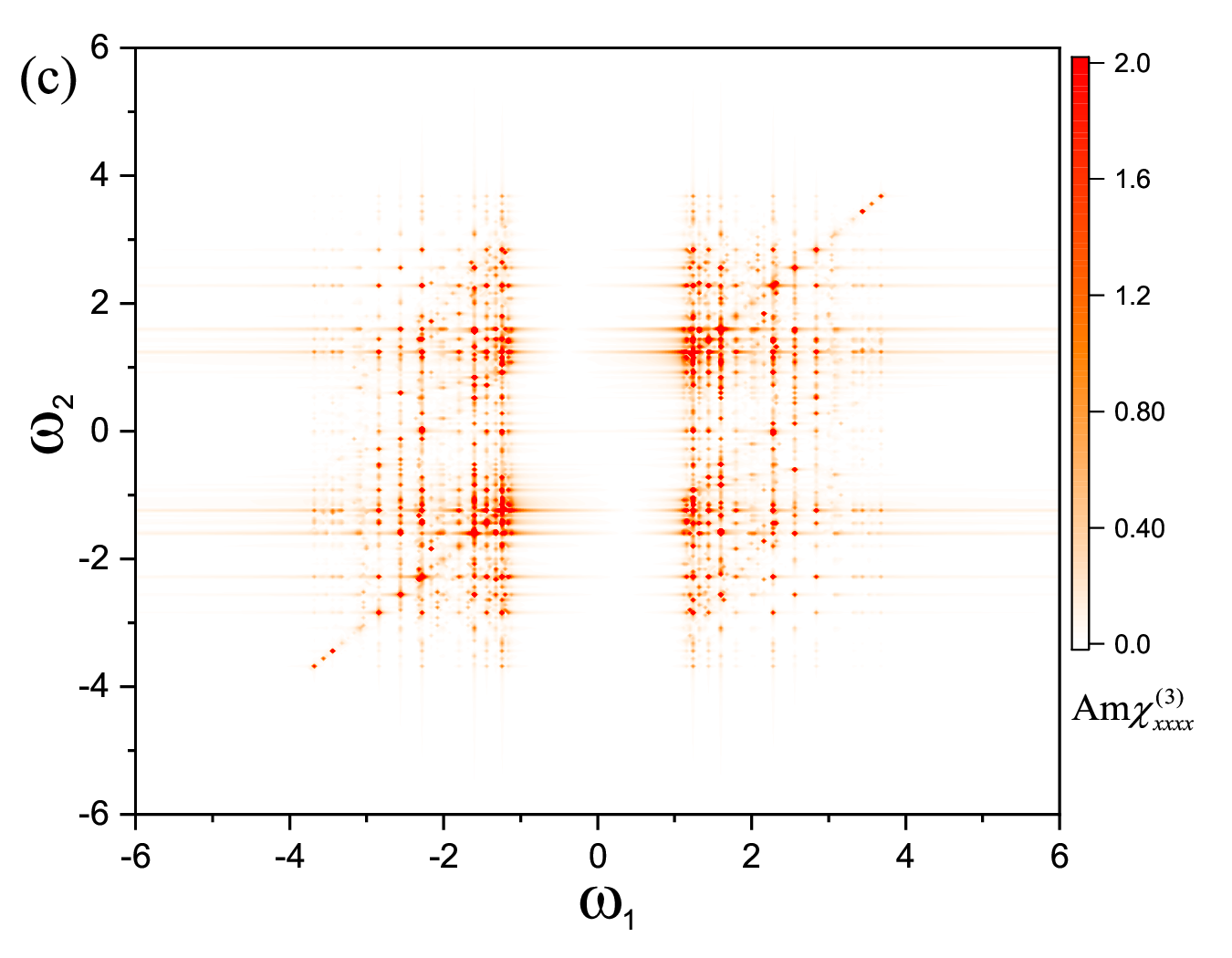} 
  	\includegraphics[angle=0, width=0.45 \columnwidth]{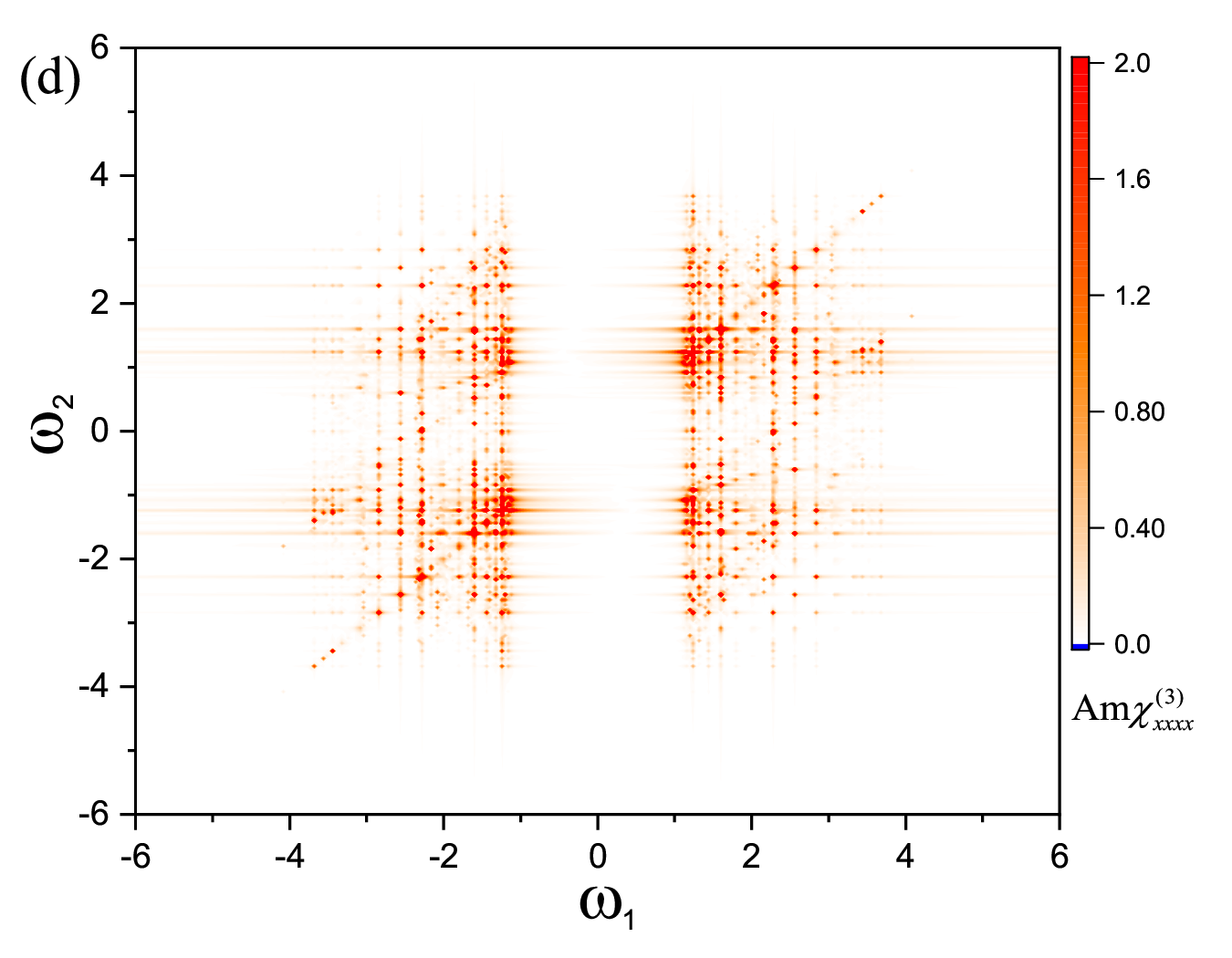}  
  	\caption {(Color online) 
   	Two-dimensional amplitude spectrum of the third-order susceptibilities 
   		$\chi^{(3)}_{xxxx}(\omega_{2},0,\omega_{1})$ for $\theta$=20\textdegree
   	with two-kink single spin-flip (a), two-kink double spin-flip (b),
  	 four-kink single spin flip (c), and four-kink double spin-flip (d), projection approximation, respectively.
  	} 		
  	\label{fig6:kink_project} 
  \end{figure}

  \subsection{Resolved components in two-dimensional coherent spectroscopy}

 The projection approximation is an effective method for analyzing the contributions
  of various excitation processes to 2DCS\cite{Watanabe-PRB-110-134443-2024,Sim-PRB-107-L100404-2023} .
  The low-energy excitations can be described by directly projecting the Hamiltonian 
  onto the two-kink or four-kink subspace in the transverse-field Ising model
  under weak confinement conditions (e.g., $ h_{z} \ll h_{x} $) \cite{Watanabe-PRB-110-134443-2024}.
  However, it is not feasible to project the TKM Hamiltonian onto two-kink or four-kink Hilbert spaces
  because the ground-state is not a simple fully-polarized ferromagnetic chain.
  To account for the contributions from various multi-kink excitations, we define the multi-kink state as 
   \begin{eqnarray}
   	\label{eq:mkink_state1}
   	|n,l_{max} \rangle=\sum_{l_{1}, \cdots, l_{n}} | j_{1},l_{1}, \cdots,j_{n},l_{n}, \cdots  \rangle   
   \end{eqnarray} 
   where $l_{max}=max\{ l_{1},l_{2},\cdots, l_{n}\}$ , $l_{i}\leq l_{max}$, and
   \begin{eqnarray}
   	\label{eq:mkink_state2}
   	| j_{1},l_{1}, \cdots,j_{n},l_{n}, \cdots  \rangle
   	=   |\cdots \uparrow \uparrow \downarrow_{j_{1}} 
   	\cdots      \downarrow_{j_{1}+l_{1}-1} \uparrow \uparrow \cdots  
   	\uparrow \uparrow \downarrow_{j_{n}} 
   	\cdots      \downarrow_{j_{n}+l_{n}-1} \uparrow \uparrow \cdots  
   	\rangle  .
   \end{eqnarray}  
     we expand the total magnetic moment 
 operator $M^{\alpha}$ of the system in terms of the number of kinks. The total magnetic moment
 is defined as
 \begin{eqnarray}
 	\label{eq:m_total}
 	M^{\alpha}=\dfrac{1}{N_{k}}\sum _{k}^{N_{k}}\sum_{i=1}^{N}m^{\alpha}_{i}e^{i\vec{k}\cdot\vec{r}}.
 \end{eqnarray}    
 The magnetic moment operator at each lattice site is projected onto the 
 subspace of multi-kink multi-spin flips 
 $m^{\alpha}_{nl_{max}i}= \mathcal{P} m^{\alpha}_{i} \mathcal{P} $, where $\mathcal{P}$ is projection operator, which is defined as
 \begin{eqnarray}
 	\label{eq:p_exp}
 	\mathcal{P}=\sum_{n=1}^{N_{max}}\sum_{l_{max}=1}^{L_{max}}  | n,l_{max} \rangle \langle n,l_{max} |.  
 \end{eqnarray}     
 By substituting this projected total magnetic moment operator into the higher-order 
 susceptibility formula Eq.(\ref{eq:nonlinear_susceptibility1}), 
 the contributions of different spinon excitation processes to the 2DCS are obtained. 
 As shown in Fig. \ref{fig6:kink_project}(a) and Fig. \ref{fig6:kink_project} (b), 
 the two-kink approximation already qualitatively reproduces the non-rephasing signals 
 in the first quadrant and the discrete rephasing signals in the fourth quadrant. 
 This indicates that the non-rephasing signal in the 2DCS
 of the TKM primarily originate from two-kink 
 single spin-flip excitations. 
 Nevertheless, the spectral weight within the region $-1<\omega_{2}<1$ is reduced
  in Fig. \ref{fig6:kink_project}(a) relative to that in Fig. \ref{fig3:tdns3}(b).  
 These missing signals are fully reproduced by the four-kink approximation,  
 as shown in Fig. \ref{fig6:kink_project}(c) and Fig. \ref{fig6:kink_project}(d) , clearly revealing that 
 these spectral weight is contributed by  multi-spinon  excitations.

\begin{figure}[htbp]
  	\centering
  	\includegraphics[angle=0, width=0.45 \columnwidth]{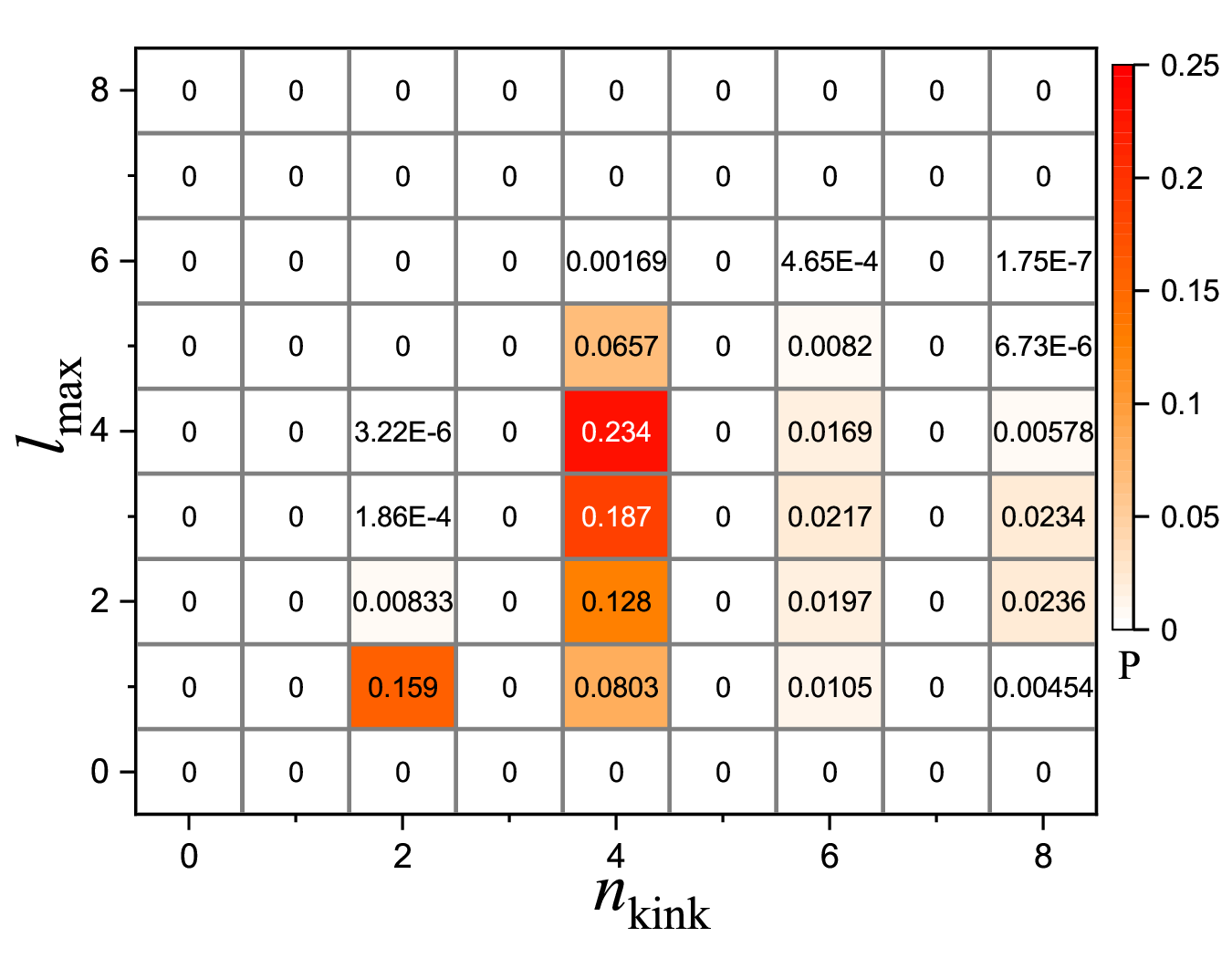} 
  	
  	\caption {(Color online) 
  		The projectd probability $|\langle n,l_{max}|\gamma\rangle|^{2}$ of multi-kink states in the excited wave function
  		$|\gamma\rangle$ for $\theta$=20\textdegree. Here, $n_\text{kink}$ denotes the number of kinks and
  	 $l_\text{max}$ represents the maximum number of flipped spins between two kinks.  
  	} 		
  	\label{fig7:probability} 
  \end{figure}

  To elucidate the  quasiparticle excitations that 
  play a dominant role in 2DCS,
 we project the excited state defined as $|\gamma\rangle = \sum_{v} \langle \Psi_{v}| M^{x} |\Psi_{n} \rangle |\Psi_{v} \rangle$
 to these multi-kink states as shown in Eq.(\ref{eq:mkink_state1}).
  The projection distributions of the excited states are shown in Figure \ref{fig7:probability}. 
  The color represents the probability defined as $|\langle n,l_\text{max}|\gamma\rangle|^{2}$, 
  where the number of spinons is given by $n_\text{kink} = 2n$.
 It is shown that two-kink single spin-flip and  four-kink two to four spin-flips
  are the most prevalent excitations, 
  which is similar to the results of transverse-field Ising model\cite{Watanabe-PRB-110-134443-2024,Coldea-Science-327-177-2010}.

    \begin{figure}[htbp]    	
  	
  	\includegraphics[angle=0, width=0.45 \columnwidth]{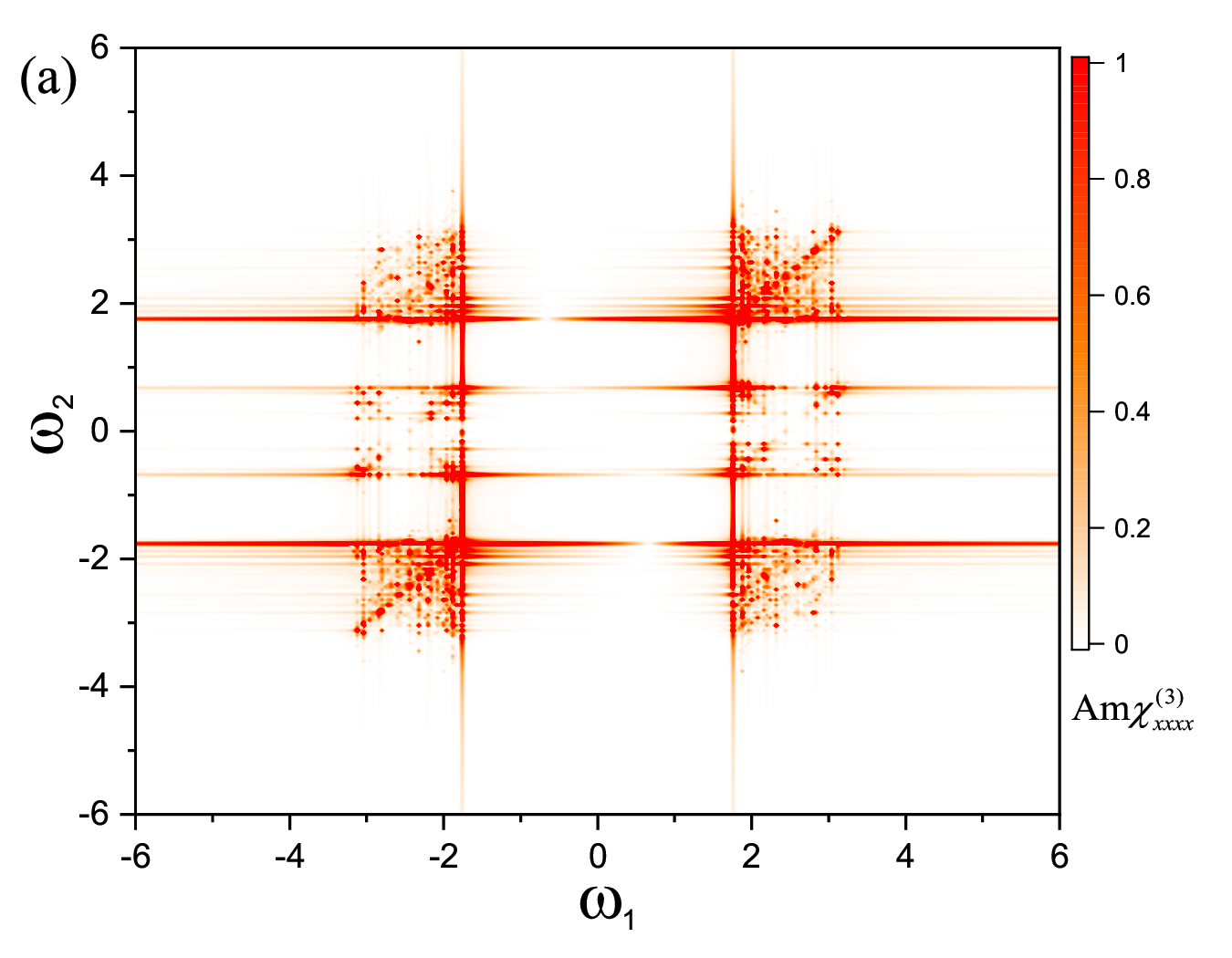}  
  	\includegraphics[angle=0, width=0.45 \columnwidth]{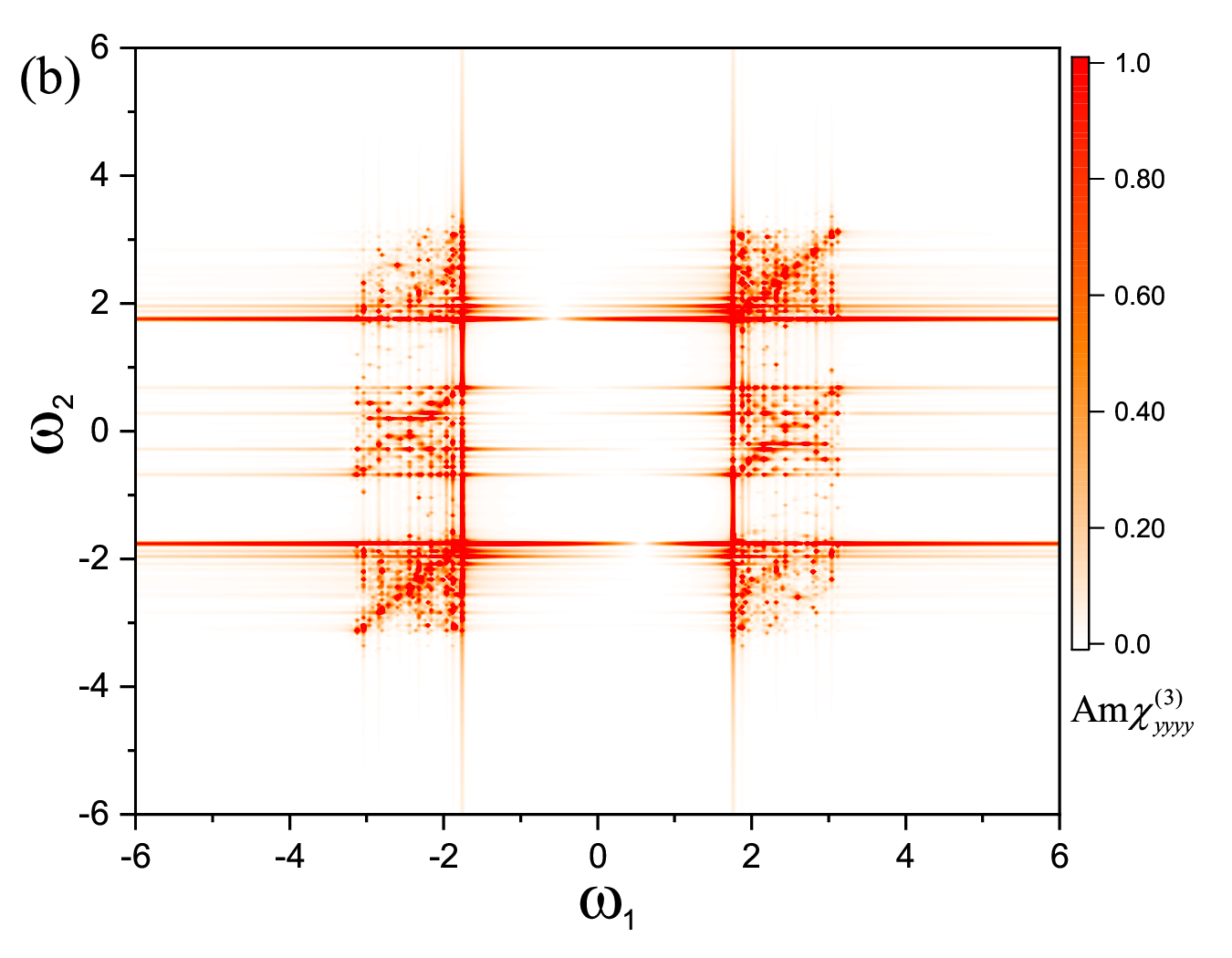}  	
  	\includegraphics[angle=0, width=0.45 \columnwidth]{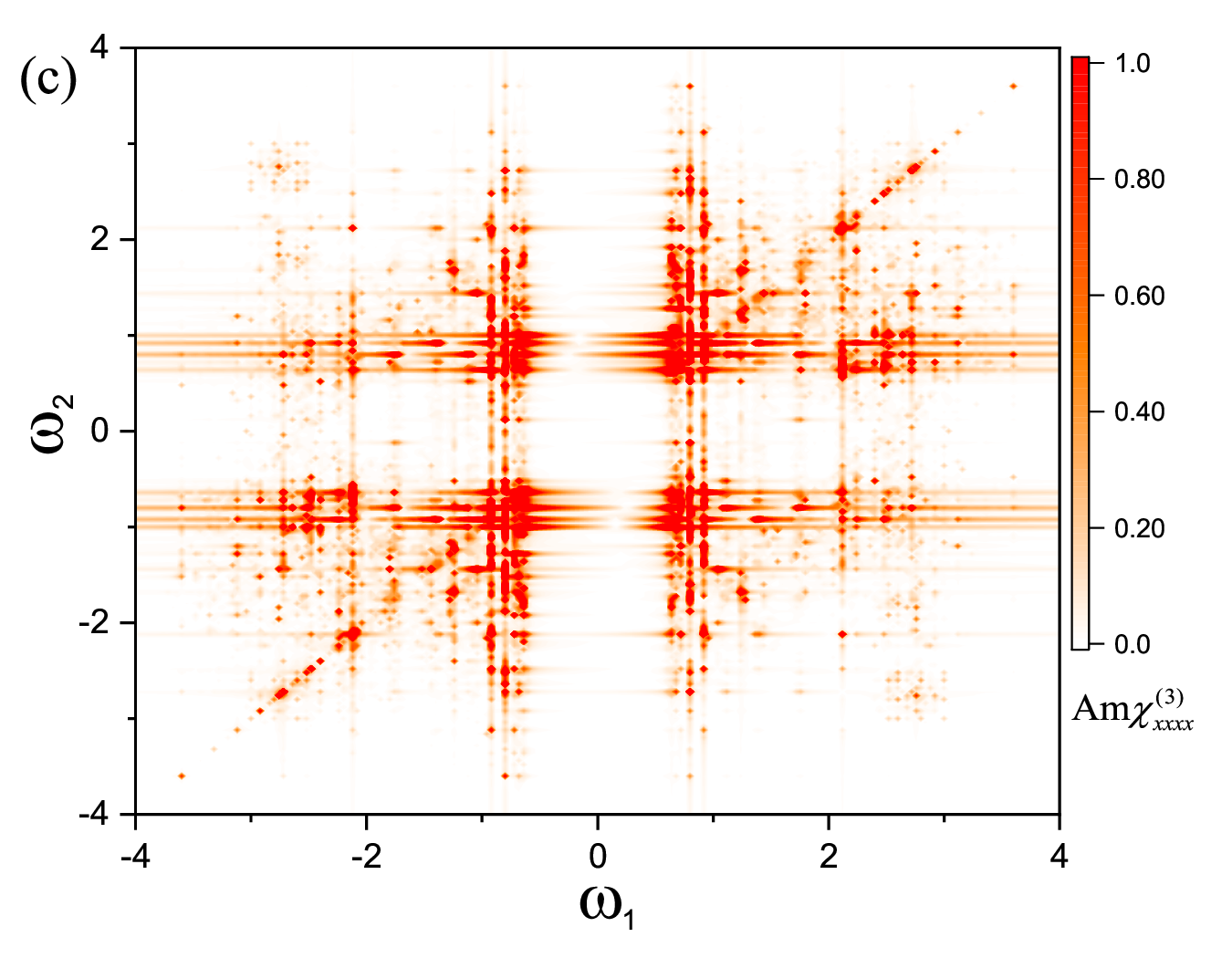}
  	\includegraphics[angle=0, width=0.45 \columnwidth]{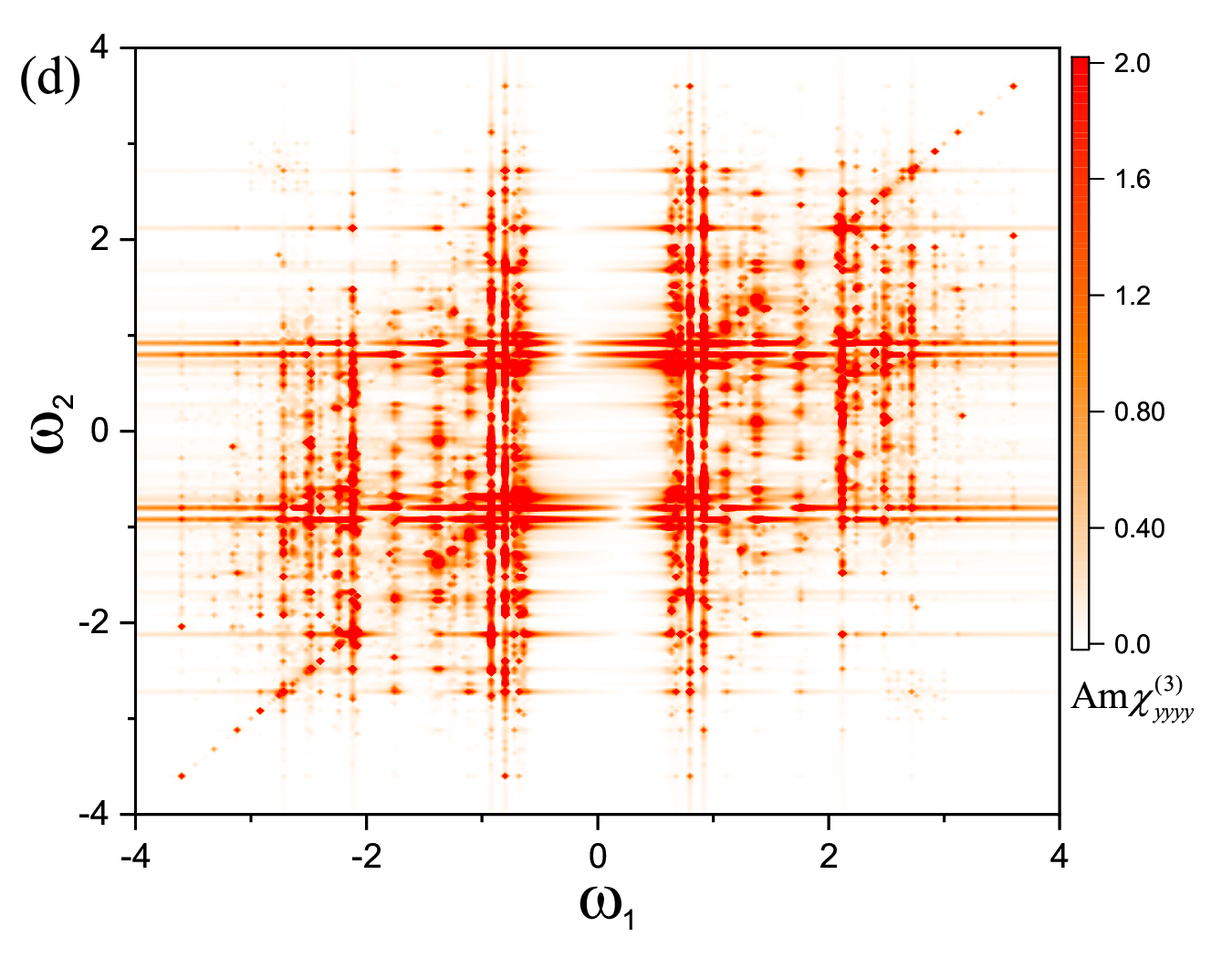} 
   	\includegraphics[angle=0, width=0.45 \columnwidth]{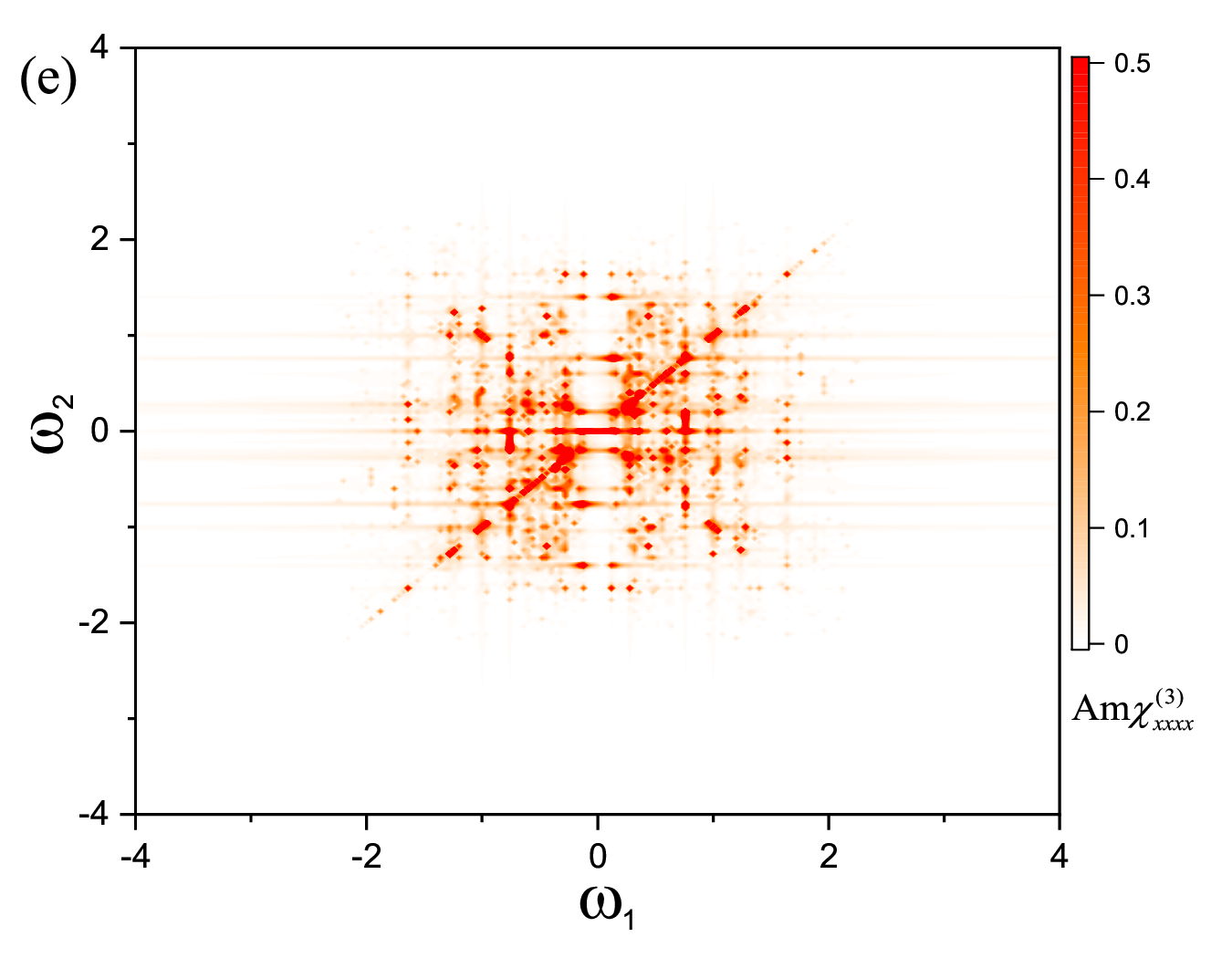}
    \includegraphics[angle=0, width=0.45 \columnwidth]{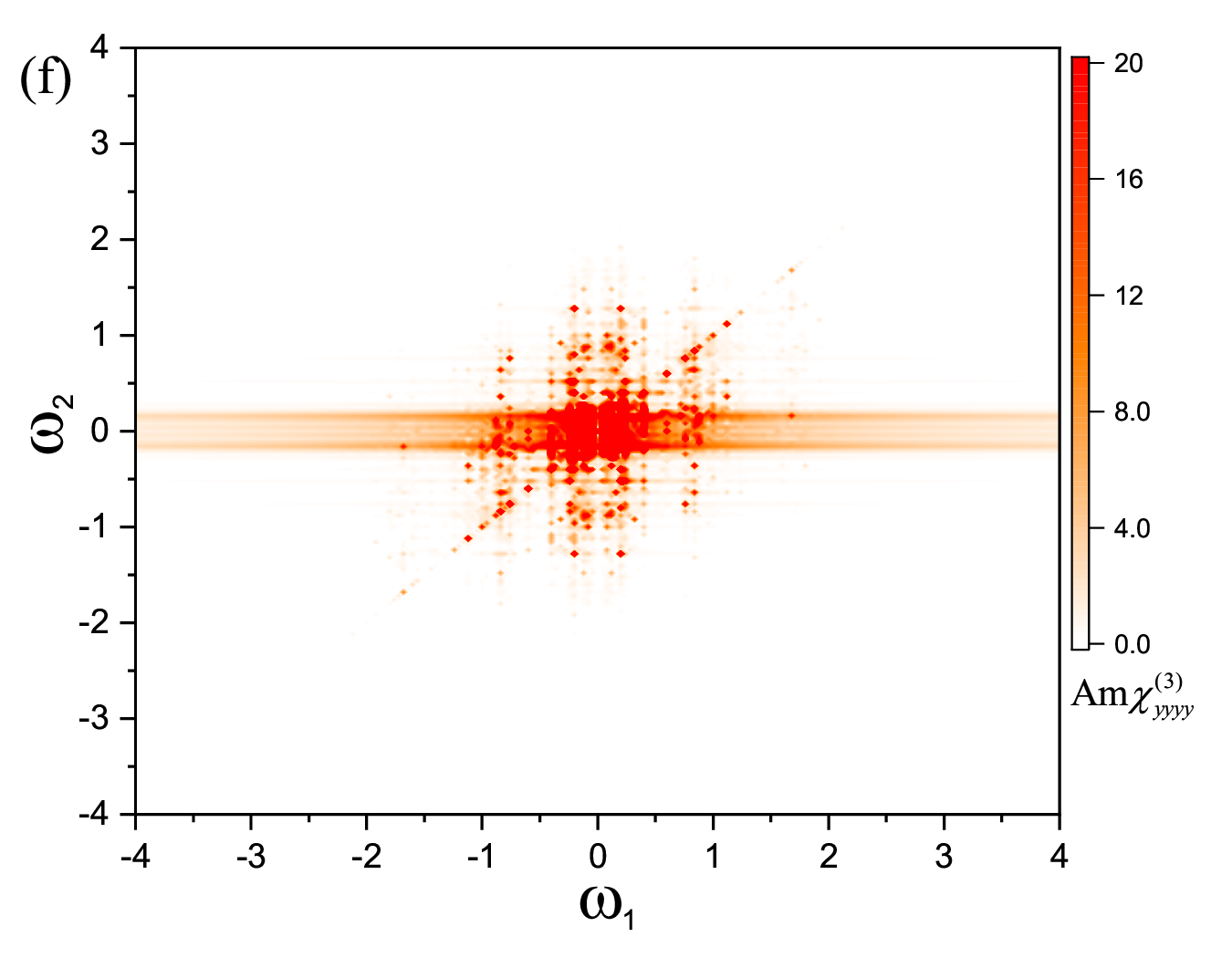}  		
  	\caption {(Color online) 
  		Two-dimensional amplitude spectrum of the third-order susceptibilities 
  		$\chi^{(3)}_{xxxx}(\omega_{2},0,\omega_{1})$ and $\chi^{(3)}_{yyyy}(\omega_{2},0,\omega_{1})$
  		with $\theta$=10\textdegree(a)(b), $\theta$=30\textdegree(c)(d),
  		$\theta$=45\textdegree(e)(f), respectively.  		
  	  The  chain length is $L=18$.}  
  	\label{fig9:TDNS_sita} 
  \end{figure}

           \begin{figure}[htbp]
  	\centering
  	\includegraphics[angle=0, width=0.45 \columnwidth]{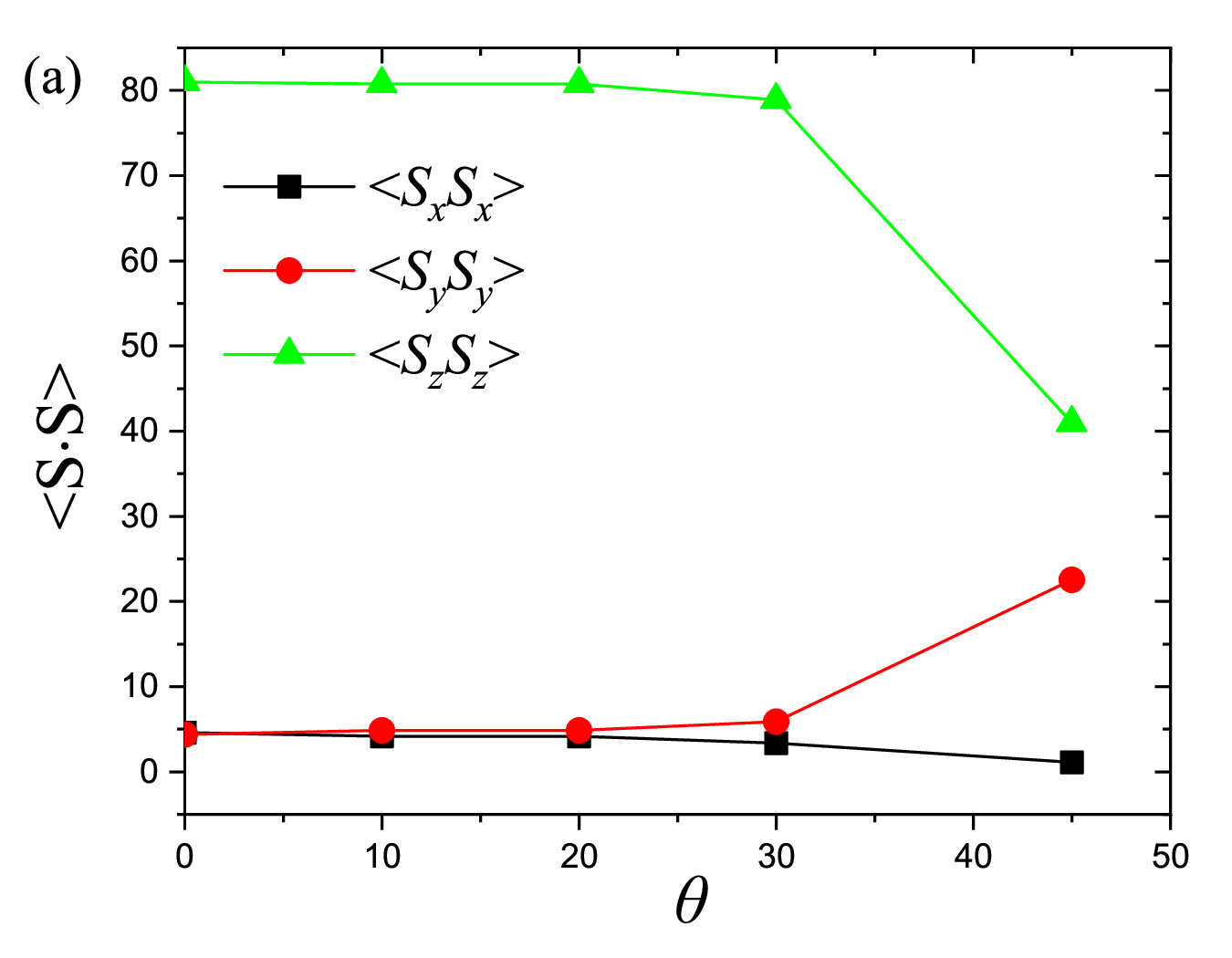} 
  	\includegraphics[angle=0, width=0.45 \columnwidth]{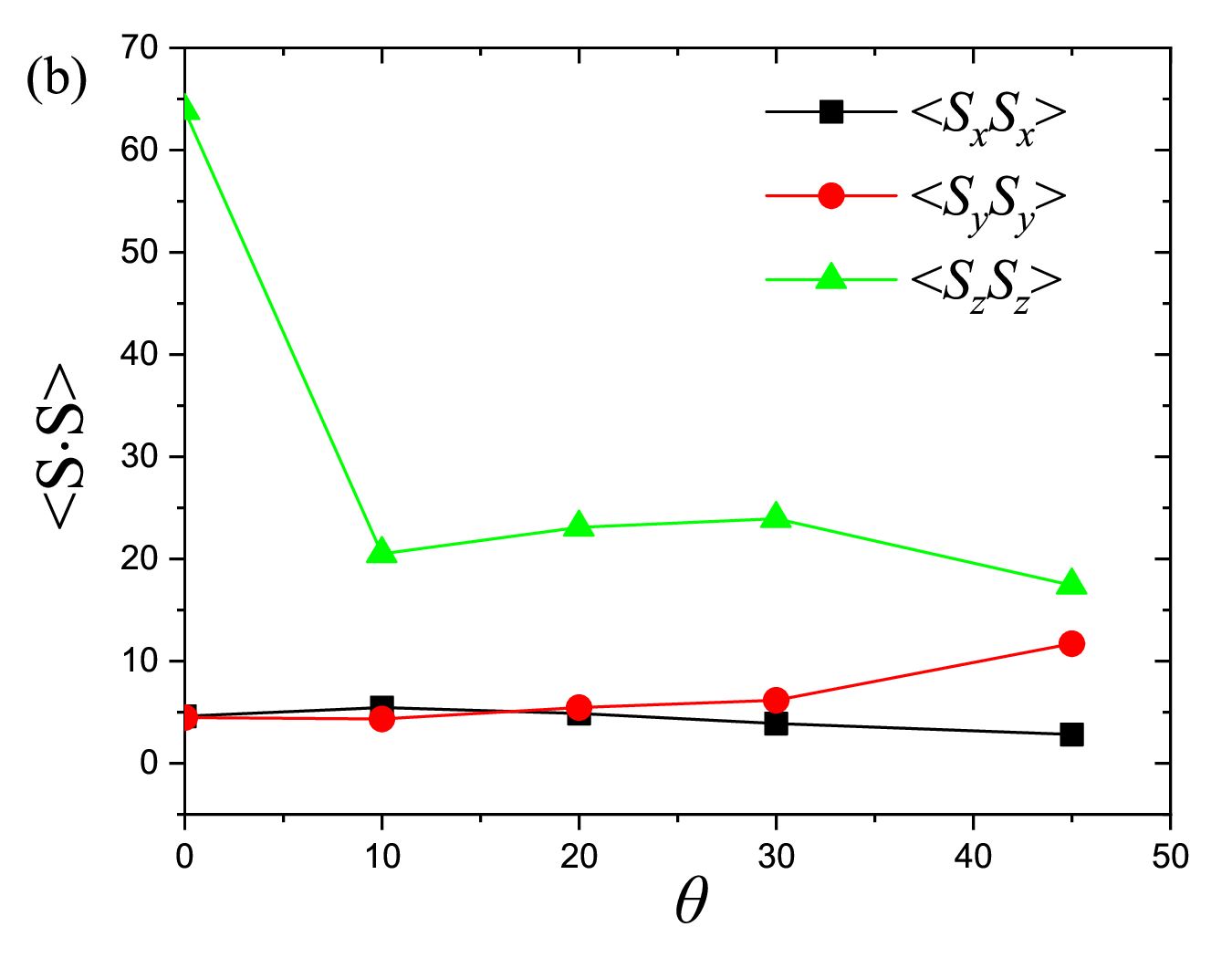} 
  	
  	\caption {(Color online) 
  		The squared total spin components of the ground state (a) and excited state $|\gamma\rangle$ (b) as
  		a function of twist angle $\theta$.
  	} 		
  	\label{fig10:magnetism} 
  \end{figure}

    \subsection{Tuning two-dimensional coherent spectroscopy with twist angle}

 To further investigate the effect of twist angle on 2DCS, 
 the spectra at several typical  angles are plotted in Fig. \ref{fig9:TDNS_sita} for comparison.
 It is shown that the energy gap 
 in the spectra significantly narrows as the twist angle increases, and
 it almost vanishes when the twist angle reaches $45$\textdegree.
 This result aligns with the results obtained by Dagotto et al \cite{Dagotto-PRB-107-104414-2023}.
In their study, the energy gap decreases as the twist angle increases
 and the excitation  gap of the system disappears at $45$\textdegree. 
  Furthermore, as the twist angle approaches $45$\textdegree, 
 the intensity of the $z$ component of dynamic spin structure factor also increases    
 markedly. 
   Correspondingly, 
 the signal intensity of $\chi^{(3),z}_{zzz}(\omega_{2},0,\omega_{1})$  component also shows a clear 
 enhancement when $\theta$ is  near $45$\textdegree in 2DCS spectra.
The spectral signal  near  $\omega_{2}=0$ in 2DCS with  $\theta=0$\textdegree
progressively  weakens and evolves into a finite number of scattered points as 
the twist angle increases. 
The reason is that spinon excitations become confined as the twist angle increases, 
leading to weakened signals primarily contributed by two-kink two-spin-flip processes.
Different from previous results,
 the rephasing signals in the fourth quadrant exhibit scattered features,
originate from finite NN $J_{x}$ and $J_{y}$ interactions \cite{Watanabe-PRB-110-134443-2024,Sim-PRB-108-134423-2023}.  
  Comparing the 2DCS in Fig. \ref{fig3:tdns3} and Fig. \ref{fig9:TDNS_sita} ,
 it is revealed that diagonal signals are  enhanced with increasing twist angle. 
  This demonstrates the feasibility of using 2DCS to probe twist-angle variations in this system.

     The squared total spin  $\langle S \cdot S \rangle=\langle \sum_{i=1}^{N}s_{i}\sum_{j=1}^{N}s_{j} \rangle$ 
     for the ground state and excited states  is plotted in Fig. \ref{fig10:magnetism} (a) and (b) respectively.
     As shown in Fig. \ref{fig10:magnetism}(a), the squared total spin  
     of the ground state remains nearly constant when the twist angle is below $30$\textdegree. 
     %
     %
     The squared total spin  of the excited state in
     Fig. \ref{fig10:magnetism}(b) indicates that approximately three to four spins are flipped,
     which is  consistent with the results presented in Fig. \ref{fig7:probability}. 
     All the results of the total spin squared collectively demonstrate that 
     the   conclusions  drawn
     for the specific twist angle  $\theta=20$\textdegree remain valid  
      across the entire $10^\circ < \theta < 30^\circ$ regime.

\section{Conclusion}

In summary, the TKM that incorporates NN exchange interactions
is modified. The finite twisting angle leads to off-diagonal spin exchange interactions 
in the laboratory coordinate system.
 Then the 2DCS  
of the quasi-one-dimensional magnetic material CoNb$_{2}$O$_{6}$ is studied  based on this model. 
The TKM exhibits sharp rephasing signals in the 2DCS
with only the $J_{z}$ exchange term\cite{Watanabe-PRB-110-134443-2024}, however,  
finite $J_{y}$ couplings lead to the spin echo signals 
appearing discrete peaks in the fourth quadrant in our results.  
 These results indicate that although the NN interactions along the $x$ and $y$ directions 
 in the local coordinate system are significantly weaker than $J_{z}$, 
 their influence on the rephasing signals in the 2DCS is also considerable.
  To investigate the contributions of different excitation processes to the 2DCS, 
we project the total magnetic moment operator onto the multi-spinon excited states.   
 Our findings indicate that 
 the diagonal non-rephasing signals and the discrete rephasing peaks in the fourth quadrant   primarily
 originate from two-spinon excitations.  
Furthermore, by comparing the spectral signals at different twisting angles and conducting 
total spin analysis, we confirmed that the above physical picture remains valid for 
twisting angles $\theta<30$\textdegree.

It should be noted that the model employed in this study does not include the 
next-nearest-neighbor superexchange interactions and inter-chain spin couplings. 
Although the inclusion of these weak interactions would improve the results, 
one could be expected that it would not qualitatively alter the present results 
since these two spin coupling strengths are weaker than the NN spin couplings at least an order in magnitude. 
On the other hand, 2DCS has been widely applied to probe magnetic, superconducting, and spin-liquid systems 
due to its exceptional ability to resolve distinct quasiparticle excitations. 
Our work establishes the fact that even weak Kitaev interactions can generate
 characteristic spectral signatures, providing a pathway to detect subtle interactions in quantum materials.


\begin{acknowledgements}

X.-L. Yu is supported by the Guangdong Basic and Applied Basic Research Foundation (Grant No. 2023A1515011852) and 
this work is also supported by MOST 2022YFA1402701.
 The calculations were  partly performed in Center for Computational Science of CASHIPS 
and Hefei Advanced Computing Center. 
 
\end{acknowledgements}

 
 \appendix

  \section{Detail of the twisted Kitaev model derivation}
 \label{app:TKmodel}

 As shown in Fig. \ref{Fig1:structure}(b), we define a laboratory coordinate system $[x,y,z]$  
 by rotating the easy magnetization axis by $\gamma=30$ around the b-axis, 
 thus the y-axis aligned with the crystal basis vector b and the z-axis 
 along the easy magnetization axis direction. Due to the presence of a twist angle, 
 local coordinate systems $[x^{\prime},y^{\prime},z^{\prime}]$ are defined based on 
 two distinct magnetic exchange directions. These local systems are obtained by rotating 
 the laboratory coordinate system by the twist angle $\theta$ about the x-axis, 
 yielding the transformations: 
  
 \begin{eqnarray}   
 	\label{appendixeq:coordinate}
 	\begin{aligned}
 	s_{i}^{x^{\prime}}&=s_{i}^{x}  \\
 	s_{i}^{y^{\prime}(\hat{n}_{1})}&=-sin(\theta)s_{i}^{z}+cos(\theta)s_{i}^{y}  \\ 
 	s_{i}^{y^{\prime}(\hat{n}_{2})}&=sin(\theta)s_{i}^{z}+cos(\theta)s_{i}^{y}  \\ 
 	s_{i}^{z^{\prime}(\hat{n}_{1})}&=cos(\theta)s_{i}^{z}+sin(\theta)s_{i}^{y}  \\  
 	s_{i}^{z^{\prime}(\hat{n}_{2})}&=cos(\theta)s_{i}^{z}-sin(\theta)s_{i}^{y}	 
 \end{aligned}
 \end{eqnarray} 
 
 Here  $\theta$ is the twist angle between the two NN  directions $\hat{n}$, where 
 $\hat{n}$ is the bond-dependent Ising interaction direction as shown 
 in Fig.\ref{Fig1:structure}(c).   
 We consider the anisotropic Heisenberg Hamiltonian in these local coordinates:
 
 \begin{eqnarray}   
 	\label{appendixeq:Heisenberg}
 	H_{\text{local}}&=&-J_{x^{\prime}}\sum_{i} s_{i}^{x^{\prime}} s_{i+1}^{x^{\prime}}
 	-J_{y^{\prime}}\sum_{i} s_{i}^{y^{\prime}} s_{i+1}^{y^{\prime}} 
 	-J_{z^{\prime}}\sum_{i} s_{i}^{z^{\prime}} s_{i+1}^{z^{\prime}}  		 
 \end{eqnarray} 
 
 where $J_{x^{\prime}}$, $J_{y^{\prime}}$ and $J_{z^{\prime}}$ are NN  magnetic 
 exchange constants in the local frames. 
 Substituting the coordinate transformations into this Hamiltonian yields the TKM Hamiltonian for NN interactions:
 
 \begin{eqnarray}   
 	\label{appendixeq:Hamiltonian1}
 	H_{\text{TKM}}&=&-J_{x^{\prime}}\sum_{i} s_{i}^{x} s_{i+1}^{x} \nonumber\\ 
 	& &-J_{y^{\prime}}\sum_{i}\left[\sin^{2}(\theta)s_{i}^{z} s_{i+1}^{z}
 	+\cos^{2}(\theta)s_{i}^{y} s_{i+1}^{y} 
 	-\dfrac{\sin(2\theta)}{2}(-1)^{i}\left(s_{i}^{z} s_{i+1}^{y}
 	+s_{i}^{y} s_{i+1}^{z} \right) \right]  \nonumber\\ 	
 	& &-J_{z^{\prime}}\sum_{i}\left[\cos^{2}(\theta)s_{i}^{z} s_{i+1}^{z}
 	+\sin^{2}(\theta)s_{i}^{y} s_{i+1}^{y} 
 	+\dfrac{\sin(2\theta)}{2}(-1)^{i}\left(s_{i}^{y} s_{i+1}^{z}
 	+s_{i}^{z} s_{i+1}^{y} \right) \right], 	 	 
 \end{eqnarray} 
 
 We write the NN magnetic interaction into matrix:

  \begin{eqnarray}   
	\label{appendixeq:Hamiltonian6}
	J&=&\left(  \begin{array}{ccc}
	-J_{x^{\prime}}	& 0 & 0 \\
	0\quad	& \quad-J_{y^{\prime}}cos^{2}(\theta)-J_{z^{\prime}}sin^{2}(\theta)\quad & \quad(-1)^{i}\left( J_{y^{\prime}}-J_{z^{\prime}} \right) \dfrac{sin(2\theta)}{2}  \\
	0\quad	& \quad(-1)^{i}\left( J_{y^{\prime}}-J_{z^{\prime}} \right) \dfrac{sin(2\theta)}{2} \quad & \quad-J_{y^{\prime}}sin^{2}(\theta)-J_{z^{\prime}}cos^{2}(\theta)
	\end{array} 	
	   \right) 
\end{eqnarray}

Compairing to the results given by referens \cite{Gallegos-PRB-109-014424-2024}, We have

  \begin{eqnarray}   
	\label{appendixeq:Hamiltonian7}
	\begin{aligned}
	J_{x^{\prime}}&= 0.57 \\ 
	-J_{y^{\prime}}cos^{2}(\theta)-J_{z^{\prime}}sin^{2}(\theta)&=-0.67	\\ 
 	\left( J_{y^{\prime}}-J_{z^{\prime}} \right) \dfrac{sin(2\theta)}{2}&=-0.56 \\ 
 	-J_{y^{\prime}}sin^{2}(\theta)-J_{z^{\prime}}cos^{2}(\theta)&=-2.48 	
 \end{aligned} 
\end{eqnarray}

 Thus the values of   exchange coupling constants are 
 $J_{x^{\prime}}=0.57$ meV, $J_{y^{\prime}}=0.51$ meV, and $J_{z^{\prime}}=2.64$ meV.

   \section{Lanczos method}
  \label{app:Lanczos}

  In the standard Lanczos method\cite{Cullum-1985}, the Krylov subspace 
  ${ |\phi_{0} \rangle, H|\phi_{0} \rangle, \cdots , H^{M-1}|\phi_{0} \rangle}$ 
  is constructed by iteratively computing matrix-vector products
   $|\phi\rangle = H|\phi_{n-1}\rangle$. Here, $|\phi_{0} \rangle$ 
   is a random initial vector. Within this Krylov subspace, 
   the Hamiltonian matrix can be reduced to a symmetric tridiagonal
    matrix of dimension $M_{L} \times M_{L}$ \cite{Avella-2013}. 
    The eigenvalues and eigenvectors of the Hamiltonian can  
    then be obtained by diagonalizing this tridiagonal matrix\cite{Cullum-1985}.  
  To preserve orthogonality among the basis vectors spanning the Krylov subspace, 
  Gram-Schmidt orthogonalization is applied to all vectors during the computation.
   This approach is termed orthogonal Lanczos method. 
   By diagonalizing the tridiagonal Hamiltonian within the Krylov subspace, 
   one obtains the ground state and several excited-state eigenvalues and eigenvectors. 
   These eigenstates are subsequently used in Eq. (\ref{eq:tdns_third11}) 
   to compute higher-order correlation functions and 2DCS.

  Furthermore, the Lanczos method has been extended to low temperature 
  and finite temperature situations\cite{Aichhorn-PRB-67-161103-2003, Jaklic-PRB-49-5065-1994},
  which can be used to  calculate the temperature-dependent magnetic and 
  thermodynamic quantities of strongly correlated systems 
  \cite{Zerec-PRB-73-245108-2006,Horsch-PRB-59-6217-1999,
  	Haule-PRB-61-2482-2000,Stephen-JAP-89-6627-2001,Jia-AnnPhys-534-2200012-2022}.
  Based on high-temperature expansions\cite{Jaklic-PRB-49-5065-1994},  
  the finite-temperature Lanczos method (FTLM) expresses the thermodynamic 
  average of an operator $A$ and the exact partition function $Z$ are given by the following equations:
  \begin{eqnarray} 
  	\label{appendix:ftlm1}
  	<A>&=&\frac{N_{st}}{ZR}\sum_{r=1}^{R}\sum_{j=0}^{M_{L}}e^{-\beta\varepsilon_{j}^{r}}
  	<r\mid\Psi_{j}^{r}><\Psi_{j}^{r}\mid A\mid r>  
  \end{eqnarray} 
  \begin{eqnarray} 
  	\label{appendix:ftlm2} 
  	<Z>&=&\frac{N_{st}}{R}\sum_{r=1}^{R}\sum_{j=0}^{M_{L}}e^{-\beta\varepsilon_{j}^{r}}
  	\mid<r\mid\Psi_{j}^{r}>\mid^{2},
  \end{eqnarray}
  where $\beta = 1/k_{B}T$, $k_{B}$ is the Boltzmann constant, 
  and $|r \rangle$ denotes the initial random state 
  $|\phi_{0} \rangle$.  $|\Psi_{j}^{r} \rangle$ 
  is the $j$-th eigenvector of the Hamiltonian 
  calculated in the Krylov 
  space  with an  initial  
  vector $|\phi_{0} \rangle = |r\rangle$. 
  Within the FTLM, a set of $R$ distinct random initial vectors $|r\rangle$ is selected. 
   $\varepsilon_{j}^{r}$ is the eigenvalue corresponding
    to the eigenstate $|\Psi_{j}^{r}\rangle$, $N_{st}$ and $M_{L}$
    denote the dimensions of the full Hamiltonian matrix and the Krylov subspace, respectively.
  To evaluate the thermodynamic statistical value $\langle A \rangle$,  
   the eigenvalues 
  $\varepsilon_{j}^{r}$ and eigenvectors $|\Psi_{j}^{r} \rangle$ 
  of the Hamiltonian are calculated via the Lanczos method using the initial 
  random vector $|r\rangle$ firstly; subsequently, the thermodynamic 
  average is obtained using Eqs. (\ref{appendix:ftlm1}) and (\ref{appendix:ftlm2}). 
  In the present work, the finite-temperature Lanczos method is employed 
  to calculate the specific heat of the TKM.

 \section{The $n$th order nonlinear susceptibility}
   \label{app:nonsus}
      
 In this Appendix, we  provide detailed calculations for the nonlinear 
 susceptibilities\cite{Kim-PRL-124-117205-2020,Armitage-PRL-122-257401-2019,Mukamel-1995}.
 According to Eq.(\ref{eq:nonlinear_susceptibility1}), the third order 
 nonlinear susceptibility is given by (we choose the unit $\hbar=1$)
  \begin{eqnarray}
 	\label{appendixA:TNS1}
 	\chi^{(3)}_{\alpha\alpha\alpha\alpha}\left( \tau_{3},\tau_{2},\tau_{1} \right)&=&
 	\frac{i^{3}}{L}\varTheta(\tau_{1})\varTheta(\tau_{2})\varTheta(\tau_{3}) 	\nonumber\\  
  & & \times \left\langle \left[ \left[ \left[ M^{\alpha}(\tau_{3}+\tau_{2}+\tau_{1}),M^{\alpha}(\tau_{2}+\tau_{1})\right],
 	M^{\alpha}(\tau_{1})\right],M^{\alpha}(0)\right] \right\rangle  	
 \end{eqnarray}  
 At two pulse limits of the three-pulse process, we get  
 \begin{eqnarray}
 	\label{appendixA:tdns_third1}
 	\chi^{(3)}_{\alpha\alpha\alpha\alpha}\left( \tau_{2},\tau_{1},0 \right)&=& -\dfrac{i}{L} 
 	\varTheta(\tau_{1})\varTheta(\tau_{2})
 	\sum_{l=1}^{8} R^{(l)}_{\alpha\alpha\alpha\alpha}\left( \tau_{2},\tau_{1},0 \right),  
 \end{eqnarray}    
 \begin{eqnarray} 
 	\label{appendixA:tdns_third2}
 	\chi^{(3)}_{\alpha\alpha\alpha\alpha}\left( \tau_{2},0,\tau_{1} \right)&=& -\dfrac{i}{L} 
 	\varTheta(\tau_{1})\varTheta(\tau_{2})
 	\sum_{l=1}^{8} R^{(l)}_{\alpha\alpha\alpha\alpha}\left( \tau_{2},0,\tau_{1} \right).  
 \end{eqnarray}   
 In Eq.(\ref{appendixA:tdns_third1}) and Eq.(\ref{appendixA:tdns_third2}), the
 high-order correlation functions are   
 \begin{eqnarray}
 	\label{appendixA:tdns_third3}
 	R^{(1)}_{\alpha\alpha\alpha\alpha}\left( t_{3},t_{2},t_{1} \right)=\langle 
 	\hat{M}^{\alpha}(t_{3}+t_{2}+t_{1}) \hat{M}^{\alpha}(t_{2}+t_{1})
 	\hat{M}^{\alpha}(t_{1})\hat{M}^{\alpha}(0)
 	\rangle,
 \end{eqnarray}  
 \begin{eqnarray}
 	\label{appendixA:tdns_third4}
 	R^{(2)}_{\alpha\alpha\alpha\alpha}\left( t_{3},t_{2},t_{1} \right)=-\langle 
 	\hat{M}^{\alpha}(t_{2}+t_{1}) \hat{M}^{\alpha}(t_{3}+t_{2}+t_{1}) 
 	\hat{M}^{\alpha}(t_{1})\hat{M}^{\alpha}(0)
 	\rangle,
 \end{eqnarray}  
 \begin{eqnarray}
 	\label{appendixA:tdns_third5}
 	R^{(3)}_{\alpha\alpha\alpha\alpha}\left( t_{3},t_{2},t_{1} \right)=-\langle 
 	\hat{M}^{\alpha}(t_{1})	 \hat{M}^{\alpha}(t_{3}+t_{2}+t_{1}) 
 	\hat{M}^{\alpha}(t_{2}+t_{1}) \hat{M}^{\alpha}(0)
 	\rangle,
 \end{eqnarray} 
 \begin{eqnarray}
 	\label{appendixA:tdns_third6}
 	R^{(4)}_{\alpha\alpha\alpha\alpha}\left( t_{3},t_{2},t_{1} \right)=\langle 
 	\hat{M}^{\alpha}(t_{1}) \hat{M}^{\alpha}(t_{2}+t_{1})	 
 	\hat{M}^{\alpha}(t_{3}+t_{2}+t_{1}) \hat{M}^{\alpha}(0)
 	\rangle,
 \end{eqnarray} 
 \begin{eqnarray}
 	\label{appendixA:tdns_third7}
 	R^{(5)}_{\alpha\alpha\alpha\alpha}\left( t_{3},t_{2},t_{1} \right)=-\langle 
 	\hat{M}^{\alpha}(0) \hat{M}^{\alpha}(t_{3}+t_{2}+t_{1})  	 
 	\hat{M}^{\alpha}(t_{2}+t_{1}) 	\hat{M}^{\alpha}(t_{1})
 	\rangle,
 \end{eqnarray} 
 \begin{eqnarray}
 	\label{appendixA:tdns_third8}
 	R^{(6)}_{\alpha\alpha\alpha\alpha}\left( t_{3},t_{2},t_{1} \right)=\langle 
 	\hat{M}^{\alpha}(0) \hat{M}^{\alpha}(t_{2}+t_{1})  	 
 	\hat{M}^{\alpha}(t_{3}+t_{2}+t_{1}) \hat{M}^{\alpha}(t_{1})
 	\rangle,
 \end{eqnarray} 
 \begin{eqnarray}
 	\label{appendixA:tdns_third9}
 	R^{(7)}_{\alpha\alpha\alpha\alpha}\left( t_{3},t_{2},t_{1} \right)=\langle 
 	\hat{M}^{\alpha}(0)\hat{M}^{\alpha}(t_{1})   	 
 	\hat{M}^{\alpha}(t_{3}+t_{2}+t_{1}) \hat{M}^{\alpha}(t_{2}+t_{1})
 	\rangle,
 \end{eqnarray} 
 \begin{eqnarray}    
 	\label{appendixA:tdns_third10}
 R^{(8)}_{\alpha\alpha\alpha\alpha}\left( t_{3},t_{2},t_{1} \right)=-\langle 
 	\hat{M}^{\alpha}(0)\hat{M}^{\alpha}(t_{1})   	 
 	\hat{M}^{\alpha}(t_{2}+t_{1})	\hat{M}^{\alpha}(t_{3}+t_{2}+t_{1})
 	\rangle.
 \end{eqnarray}

 It is obvious that: 
  $R^{(5)}_{\alpha\alpha\alpha\alpha}=-\left[ R^{(4)}_{\alpha\alpha\alpha\alpha} \right]^{\ast}$,
  $R^{(6)}_{\alpha\alpha\alpha\alpha}=-\left[ R^{(3)}_{\alpha\alpha\alpha\alpha} \right]^{\ast}$,
  $R^{(7)}_{\alpha\alpha\alpha\alpha}=-\left[ R^{(2)}_{\alpha\alpha\alpha\alpha} \right]^{\ast}$,
  $R^{(8)}_{\alpha\alpha\alpha\alpha}=-\left[ R^{(1)}_{\alpha\alpha\alpha\alpha} \right]^{\ast}$. 
 We decompose the correlation 
 function by inserting the resolution of identity in Krylov subspace. 
 The third order nonlinear susceptibility is written as : 
   \begin{eqnarray} 
 	\label{appendixA:TNS2}
 	\begin{split}
 	&\chi^{(3)}_{\alpha\alpha\alpha\alpha}\left( \tau_{3},\tau_{2},\tau_{1} \right)=
 	 \frac{i^{3}}{L}\varTheta(\tau_{1})\varTheta(\tau_{2})\varTheta(\tau_{3}) 
 	 \dfrac{1}{Z}\sum_{n}e^{-\beta \varepsilon_{n}}\sum_{pqv=0}^{M_{krylov}}\left[ C_{R}+iC_{I}\right] \\
 	 & \times \left\lbrace 
 	 e^{\frac{i}{\hbar}(\varepsilon_{n}-\varepsilon_{v})\tau_{1}} 
 	 e^{\frac{i}{\hbar}(\varepsilon_{n}-\varepsilon_{q})\tau_{2}}  
 	 e^{\frac{i}{\hbar}(\varepsilon_{n}-\varepsilon_{p})\tau_{3}}
 	-e^{\frac{i}{\hbar}(\varepsilon_{n}-\varepsilon_{v})\tau_{1}} 
 	 e^{\frac{i}{\hbar}(\varepsilon_{n}-\varepsilon_{q})\tau_{2}}  
 	 e^{\frac{i}{\hbar}(\varepsilon_{p}-\varepsilon_{q})\tau_{3}} \right. \\
  	& -e^{\frac{i}{\hbar}(\varepsilon_{n}-\varepsilon_{v})\tau_{1}} 
 	 e^{\frac{i}{\hbar}(\varepsilon_{p}-\varepsilon_{v})\tau_{2}}  
 	 e^{\frac{i}{\hbar}(\varepsilon_{p}-\varepsilon_{q})\tau_{3}}
 	 +e^{\frac{i}{\hbar}(\varepsilon_{n}-\varepsilon_{v})\tau_{1}} 
 	 e^{\frac{i}{\hbar}(\varepsilon_{p}-\varepsilon_{v})\tau_{2}}  
 	 e^{\frac{i}{\hbar}(\varepsilon_{q}-\varepsilon_{v})\tau_{3}}   \left.   \right\rbrace \\ 
  	& + \frac{i^{3}}{L}\varTheta(\tau_{1})\varTheta(\tau_{2})\varTheta(\tau_{3}) 
  	 \dfrac{1}{Z}\sum_{n}e^{-\beta \varepsilon_{n}}\sum_{pqv=0}^{M_{krylov}}\left[ C_{R}-iC_{I}\right] \\
 	  & \times \left\lbrace 	 	 	 
  	  -e^{\frac{i}{\hbar}(\varepsilon_{v}-\varepsilon_{n})\tau_{1}} 
  e^{\frac{i}{\hbar}(\varepsilon_{v}-\varepsilon_{p})\tau_{2}}  
  e^{\frac{i}{\hbar}(\varepsilon_{v}-\varepsilon_{q})\tau_{3}}
  +e^{\frac{i}{\hbar}(\varepsilon_{v}-\varepsilon_{n})\tau_{1}} 
  e^{\frac{i}{\hbar}(\varepsilon_{v}-\varepsilon_{p})\tau_{2}}  
  e^{\frac{i}{\hbar}(\varepsilon_{q}-\varepsilon_{p})\tau_{3}} \right. \\ 
  & +e^{\frac{i}{\hbar}(\varepsilon_{v}-\varepsilon_{n})\tau_{1}} 
 e^{\frac{i}{\hbar}(\varepsilon_{q}-\varepsilon_{n})\tau_{2}}  
 e^{\frac{i}{\hbar}(\varepsilon_{q}-\varepsilon_{p})\tau_{3}}
 -e^{\frac{i}{\hbar}(\varepsilon_{v}-\varepsilon_{n})\tau_{1}} 
 e^{\frac{i}{\hbar}(\varepsilon_{q}-\varepsilon_{n})\tau_{2}}  
 e^{\frac{i}{\hbar}(\varepsilon_{p}-\varepsilon_{n})\tau_{3}}  	  
     \left.   \right\rbrace, 	   
 	\end{split}	 	 
 \end{eqnarray} 
  where $C_{R}+iC_{I}=\langle \Psi_{n}|M^{\alpha}|\Psi_{p}\rangle\langle \Psi_{p}|M^{\alpha}|
  \Psi_{q}\rangle\langle \Psi_{q}|M^{\alpha}|\Psi_{v}\rangle\langle \Psi_{v}|M^{\alpha}|\Psi_{n}\rangle$.
 By considering $\tau_{2}=0$ and $\tau_{3}\longrightarrow \tau_{2}$,
 we get the two pulse limit formula  
    \begin{eqnarray}
 	\label{appendixA:TNS3} 
 	\begin{split}
	&\chi^{(3)}_{\alpha\alpha\alpha\alpha}\left( \tau_{2},0,\tau_{1} \right)=
	\frac{i^{3}}{L}\varTheta(\tau_{1})\varTheta(\tau_{2})
	\dfrac{1}{Z}\sum_{n}e^{-\beta \varepsilon_{n}}\sum_{pqv=0}^{M_{krylov}}\left[ C_{R}+iC_{I}\right] \\
	& \times \left\lbrace 
	e^{\frac{i}{\hbar}(\varepsilon_{n}-\varepsilon_{v})\tau_{1}}  
	e^{\frac{i}{\hbar}(\varepsilon_{n}-\varepsilon_{p})\tau_{2}}
	-e^{\frac{i}{\hbar}(\varepsilon_{n}-\varepsilon_{v})\tau_{1}} 
	e^{\frac{i}{\hbar}(\varepsilon_{p}-\varepsilon_{q})\tau_{2}} \right. \\
	& -e^{\frac{i}{\hbar}(\varepsilon_{n}-\varepsilon_{v})\tau_{1}}
	e^{\frac{i}{\hbar}(\varepsilon_{p}-\varepsilon_{q})\tau_{2}}
	+e^{\frac{i}{\hbar}(\varepsilon_{n}-\varepsilon_{v})\tau_{1}}
	e^{\frac{i}{\hbar}(\varepsilon_{q}-\varepsilon_{v})\tau_{2}}   \left.   \right\rbrace \\ 
	& + \frac{i^{3}}{L}\varTheta(\tau_{1})\varTheta(\tau_{2}) 
	\dfrac{1}{Z}\sum_{n}e^{-\beta \varepsilon_{n}}\sum_{pqv=0}^{M_{krylov}}\left[ C_{R}-iC_{I}\right] \\
	& \times \left\lbrace 	 	 	 
	-e^{\frac{i}{\hbar}(\varepsilon_{v}-\varepsilon_{n})\tau_{1}} 
	e^{\frac{i}{\hbar}(\varepsilon_{v}-\varepsilon_{q})\tau_{2}}
	+e^{\frac{i}{\hbar}(\varepsilon_{v}-\varepsilon_{n})\tau_{1}} 
	e^{\frac{i}{\hbar}(\varepsilon_{q}-\varepsilon_{p})\tau_{2}} \right. \\ 
	& +e^{\frac{i}{\hbar}(\varepsilon_{v}-\varepsilon_{n})\tau_{1}} 
	e^{\frac{i}{\hbar}(\varepsilon_{q}-\varepsilon_{p})\tau_{2}}
	-e^{\frac{i}{\hbar}(\varepsilon_{v}-\varepsilon_{n})\tau_{1}} 
	e^{\frac{i}{\hbar}(\varepsilon_{p}-\varepsilon_{n})\tau_{2}}  	  
	\left.   \right\rbrace.	   
\end{split}	 
 \end{eqnarray}
  
  Applying a Fourier transform, the space-time Fourier transform of $\chi$ is expressed as follows 
 \begin{eqnarray} 
 \label{appendixA:TNS4}  
 		 \chi^{(3)}_{\alpha\alpha\alpha\alpha}\left( \omega_{2},0,\omega_{1} \right)&=&
 		\dfrac{i}{ZL}\sum_{n}e^{-\beta \varepsilon_{n}}\sum_{pqv=0}^{M_{krylov}}
 		\left[ C_{R}+iC_{I}\right] F_{201}(\omega_{1},\omega_{2},\varepsilon_{n} 
 		,\varepsilon_{p},\varepsilon_{q},\varepsilon_{v})  \nonumber\\ 		
 	 & &	+\dfrac{i}{ZL}\sum_{n}e^{-\beta \varepsilon_{n}}\sum_{pqv=0}^{M_{krylov}}
 		\left[ C_{R}-iC_{I}\right] F_{201}(\omega_{1},\omega_{2},-\varepsilon_{n} 
 		,-\varepsilon_{p},-\varepsilon_{q},-\varepsilon_{v}),  
 \end{eqnarray} 
 where
  \begin{eqnarray}
 	\label{appendixA:TNS5} 
 	\begin{split} 
 &F_{201}(\omega_{1},\omega_{2},\varepsilon_{n} 
 ,\varepsilon_{p},\varepsilon_{q},\varepsilon_{v})= \\
 &\left\lbrace 
 \dfrac{1}{(\omega_{1}+\varepsilon_{n}-\varepsilon_{v}+i0^{+})
 (\omega_{2}+\varepsilon_{n}-\varepsilon_{p}+i0^{+})} \right. \\    
 &- \dfrac{1}{(\omega_{1}+\varepsilon_{n}-\varepsilon_{v}+i0^{+})
 	(\omega_{2}+\varepsilon_{p}-\varepsilon_{q}+i0^{+})}  \\      
 & - \dfrac{1}{(\omega_{1}+\varepsilon_{n}-\varepsilon_{v}+i0^{+})
 	(\omega_{2}+\varepsilon_{p}-\varepsilon_{q}+i0^{+})}     \\ 
 &\left. + \dfrac{1}{(\omega_{1}+\varepsilon_{n}-\varepsilon_{v}+i0^{+})
 	(\omega_{2}+\varepsilon_{q}-\varepsilon_{v}+i0^{+})} 
    \right\rbrace.       
 	\end{split}       
 \end{eqnarray} 
For $\tau_{1}=0$, $\tau_{3}\longrightarrow \tau_{2}$,
$\tau_{2}\longrightarrow \tau_{1}$ case, we obtain the formula 
as similiar as Eq.(\ref{appendixA:TNS4}):
 \begin{eqnarray} 
	\label{appendixA:TNS6}  
	\chi^{(3)}_{\alpha\alpha\alpha\alpha}\left( \omega_{2},\omega_{1},0 \right)&=& 
	\dfrac{i}{ZL}\sum_{n}e^{-\beta \varepsilon_{n}}\sum_{pqv=0}^{M_{krylov}}
	\left[ C_{R}+iC_{I}\right] F_{210}(\omega_{1},\omega_{2},\varepsilon_{n} 
	,\varepsilon_{p},\varepsilon_{q},\varepsilon_{v}) \nonumber\\ 
   &+&\dfrac{i}{ZL}\sum_{n}e^{-\beta \varepsilon_{n}}\sum_{pqv=0}^{M_{krylov}}
	\left[ C_{R}-iC_{I}\right] F_{210}(\omega_{1},\omega_{2},-\varepsilon_{n} 
	,-\varepsilon_{p},-\varepsilon_{q},-\varepsilon_{v}),  
\end{eqnarray}
where $F_{210}(\omega_{1},\omega_{2},\varepsilon_{n} 
,\varepsilon_{p},\varepsilon_{q},\varepsilon_{v})$ is written as
  \begin{eqnarray}
	\label{appendixA:TNS7} 
	\begin{split} 
		&F_{210}(\omega_{1},\omega_{2},\varepsilon_{n} 
		,\varepsilon_{p},\varepsilon_{q},\varepsilon_{v})= \\
		&\left\lbrace 
		\dfrac{1}{(\omega_{1}+\varepsilon_{n}-\varepsilon_{q}+i0^{+})
			(\omega_{2}+\varepsilon_{n}-\varepsilon_{p}+i0^{+})} \right. \\     
		&- \dfrac{1}{(\omega_{1}+\varepsilon_{n}-\varepsilon_{q}+i0^{+})
			(\omega_{2}+\varepsilon_{p}-\varepsilon_{q}+i0^{+})}  \\      
		& - \dfrac{1}{(\omega_{1}+\varepsilon_{p}-\varepsilon_{v}+i0^{+})
			(\omega_{2}+\varepsilon_{p}-\varepsilon_{q}+i0^{+})}  \\   
		&\left.+ \dfrac{1}{(\omega_{1}+\varepsilon_{p}-\varepsilon_{v}+i0^{+})
			(\omega_{2}+\varepsilon_{q}-\varepsilon_{v}+i0^{+})} 
		\right\rbrace.       
	\end{split}       
\end{eqnarray}

In the calculation of the correlation function
$C_{R}+iC_{I}$ 
based on the orthogonal Lanczos method, we utilize $M_{L}=2048$ eigenstates. 
This ensures the maximum excitation energy considered sufficiently 
covers the spectral (energy) range of interest.
The temperature parameter is set to $T=1.0$ K. 
To reduce the number of channels in Liouville space, only correlation function terms 
with magnitudes exceeding 0.001 are included in the summation.
Since the Fourier transform operates only over the positive time axis and the 
laser pulses are in the impulsive limit, the third-order susceptibility convolves with 
$\left[ \delta(\omega_{1}-\varepsilon_{n}-\varepsilon_{v})+\mathcal{P}i/(\omega_{1}-\varepsilon_{n}-\varepsilon_{v})\right] 
\left[\delta(\omega_{2}-\varepsilon_{n}-\varepsilon_{p})+\mathcal{P}i/(\omega_{2}-\varepsilon_{n}-\varepsilon_{p})\right]$ ,
where $\mathcal{P}$ denotes the principal value integral. 
The principal value part introduces a weak $1/\omega$ broadening along both frequency axes. 
In our numerical implementation, the broadening width for the delta functions is set to $0.002$, 
resulting in sharp features in the calculated 2DCS signal.

         \begin{figure}[htbp]    	
   	
   	\includegraphics[angle=0, width=0.42 \columnwidth]{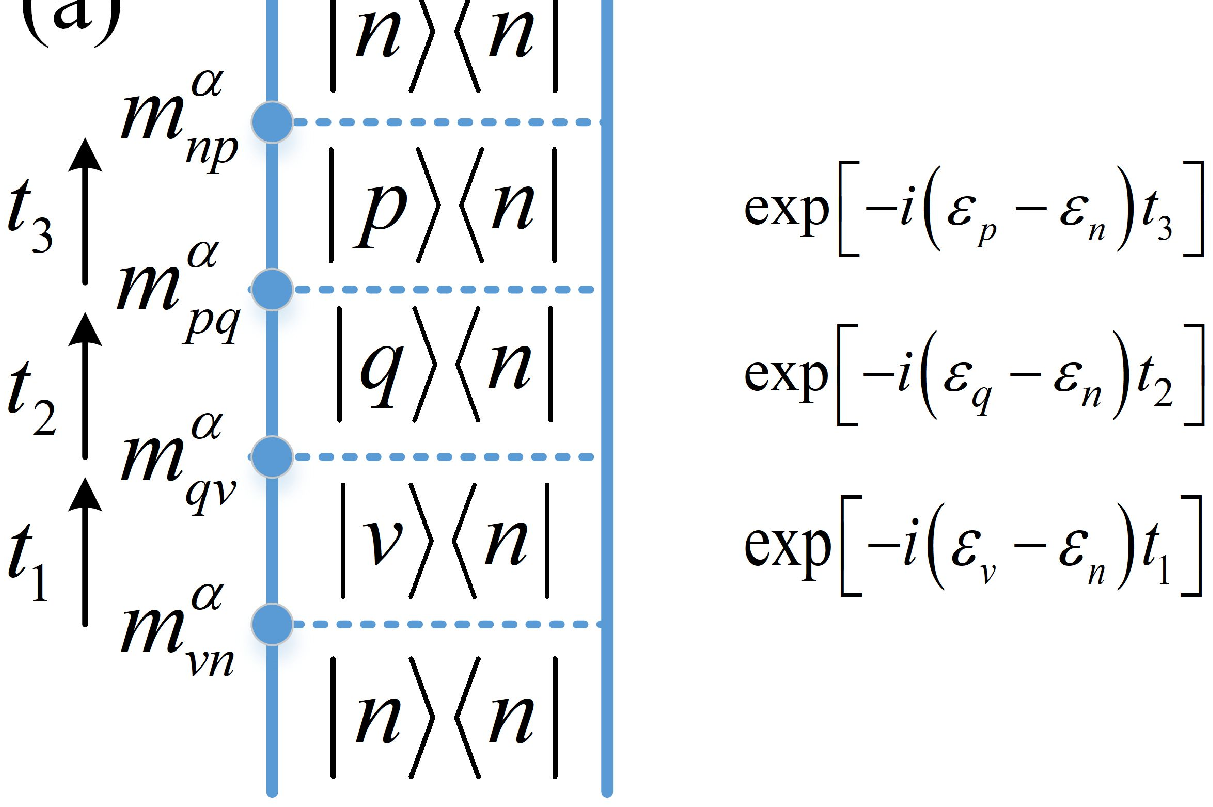}  
   	\includegraphics[angle=0, width=0.42 \columnwidth]{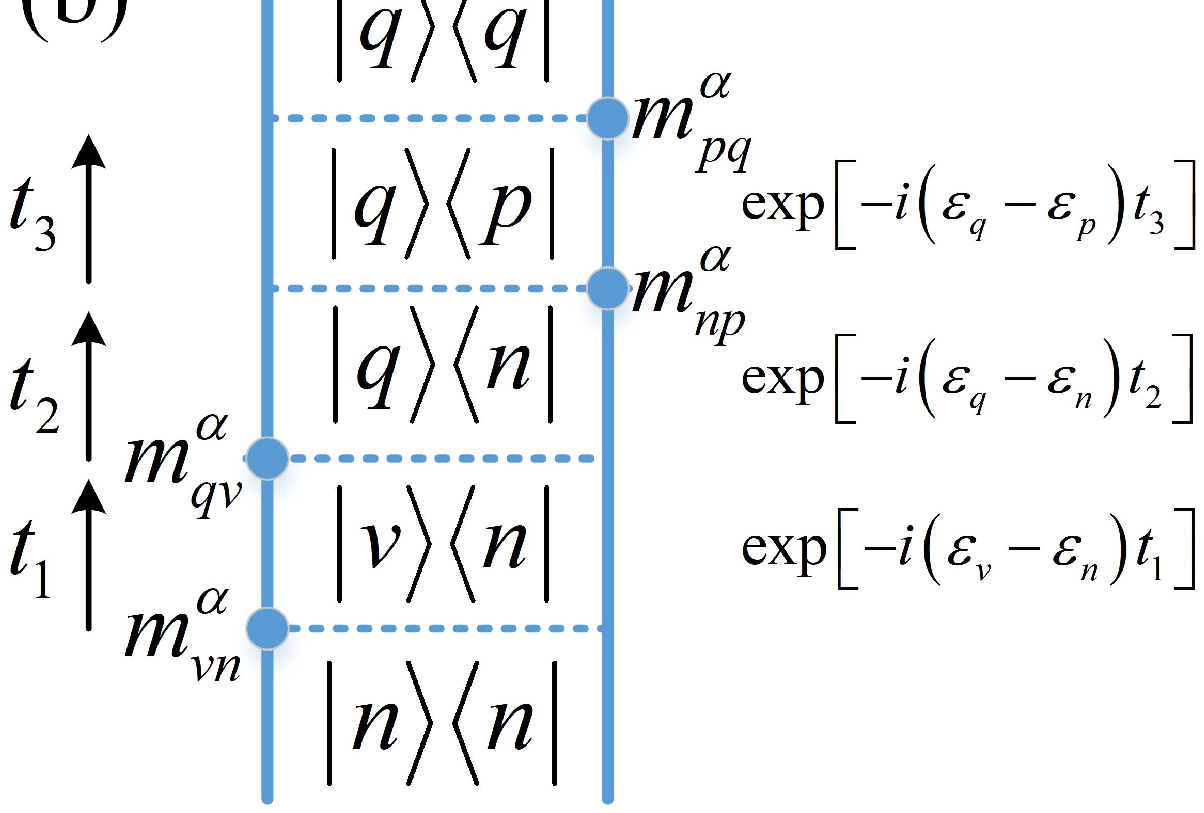} 
   	\includegraphics[angle=0, width=0.42 \columnwidth]{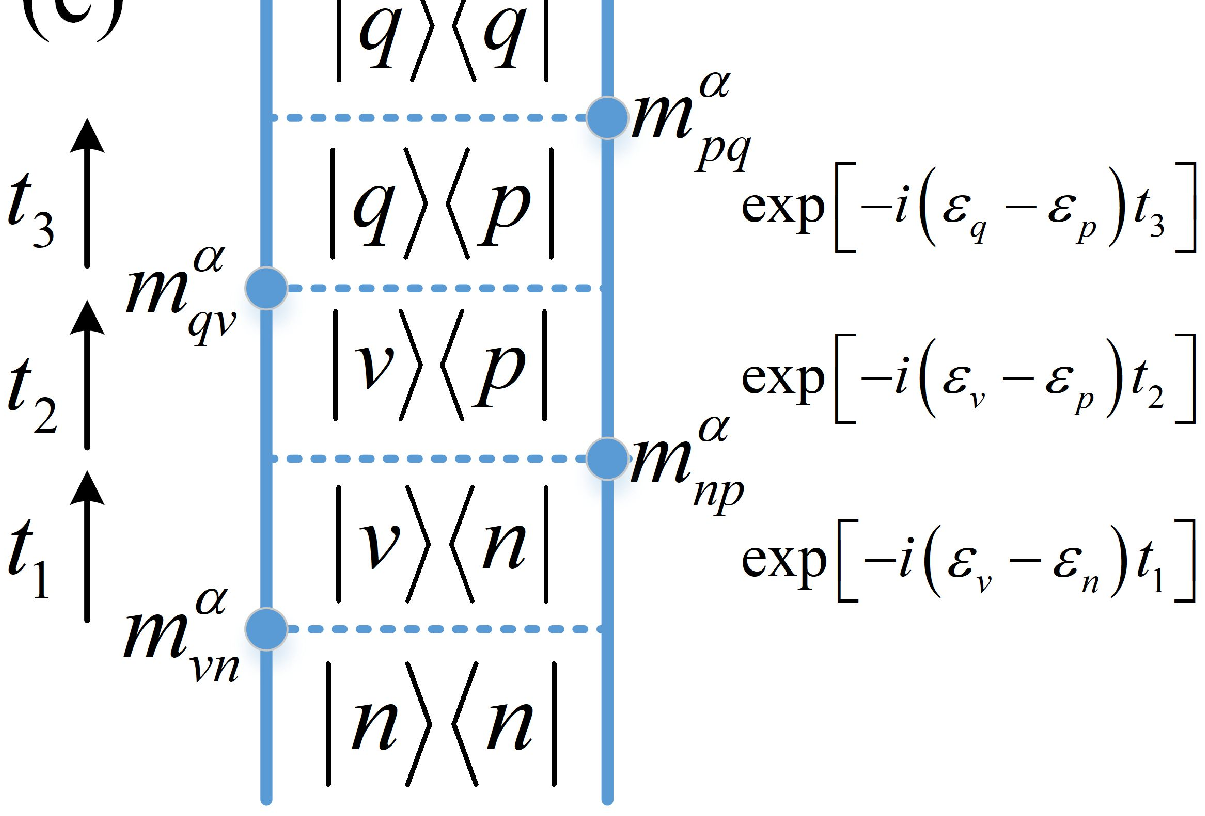}  
   	\includegraphics[angle=0, width=0.42 \columnwidth]{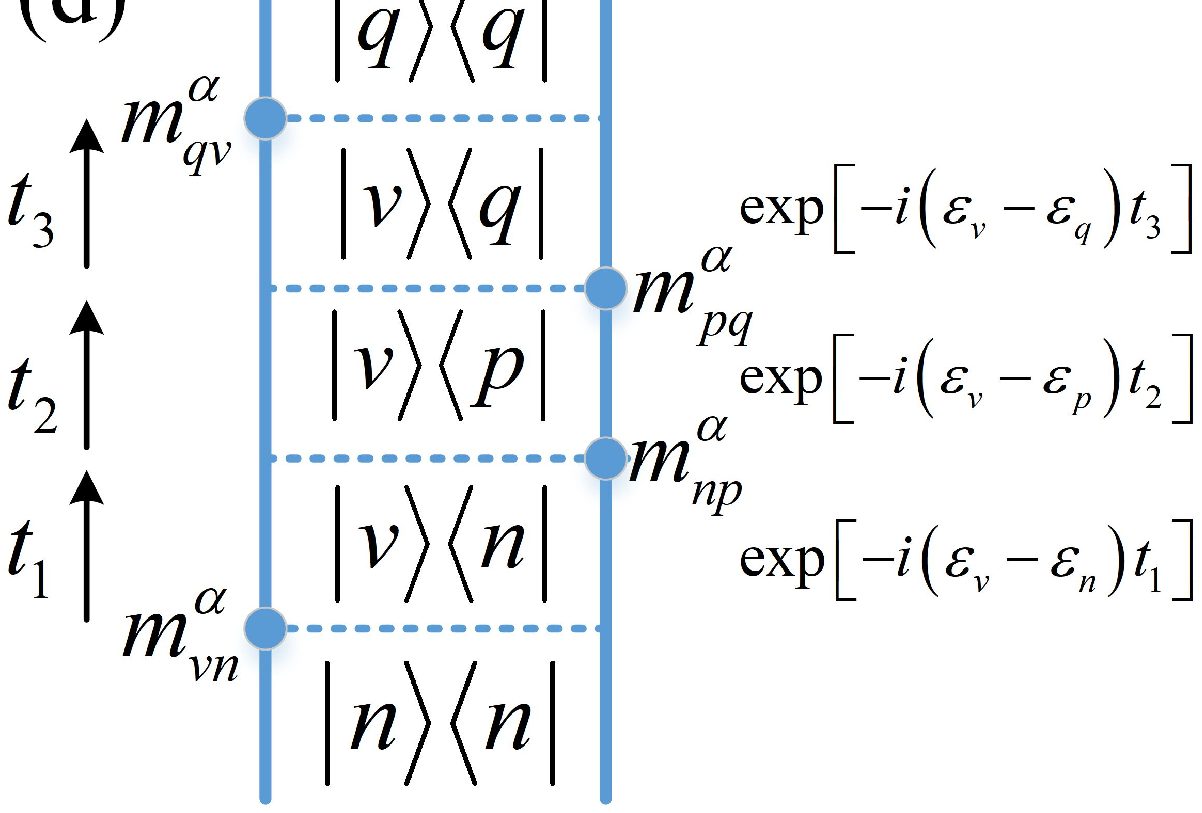}	 
   	\caption {(Color online) 
   		Double-sided Feynman diagrams for 
   		$R_{\alpha\alpha\alpha\alpha}^{(1)}$ (a),
   		$R_{\alpha\alpha\alpha\alpha}^{(2)}$ (b),
   		$R_{\alpha\alpha\alpha\alpha}^{(3)}$ (c) and
   		$R_{\alpha\alpha\alpha\alpha}^{(4)}$ (d) processes
   		that contributed to the third order coherent spectroscopy $\chi_{xxxx}^{(3)}(\omega_{2},0,\omega_{1})$.
   		Time evolves from bottom to top and dots represent bra or ket operations on the
   		density matrix. Here $|n \rangle$ is the ground state $|\Psi_{n} \rangle$ .
   		$|v \rangle$ $|p \rangle$ and $|q \rangle$ denote excited states 
   		$|\Psi_{v} \rangle$ $|\Psi_{p} \rangle$ and $|\Psi_{q} \rangle$  of the Hamiltonian respectively
   		.} 
   	\label{fig11:Feynman_diagram} 
   \end{figure}

   Fig.\ref{fig11:Feynman_diagram} displays Feynman diagrams 
   for the four third-order excitation pathways 
   $R_{\alpha \alpha \alpha\alpha}^{(1)}$ to  $R_{\alpha \alpha \alpha\alpha}^{(4)}$. 
   in two-dimensional coherent spectroscopy (2DCS).
   When $t_{2}=0$, $|\Psi_{q}\rangle=|\Psi_{n}\rangle$, $|\Psi_{v}\rangle=|\Psi_{p}\rangle$,
   the pathways represented in Figs. \ref{fig11:Feynman_diagram}(b) and (c) 
   contain the phase factor $\text{exp}[-i(\varepsilon_{v}-\varepsilon_{n})(t_{1}-t_{3})]$. 
   This functional form indicates that these excitation processes generate the rephasing signal.

     \begin{figure}[htbp]    	
 	
 	\includegraphics[angle=0, width=0.45 \columnwidth]{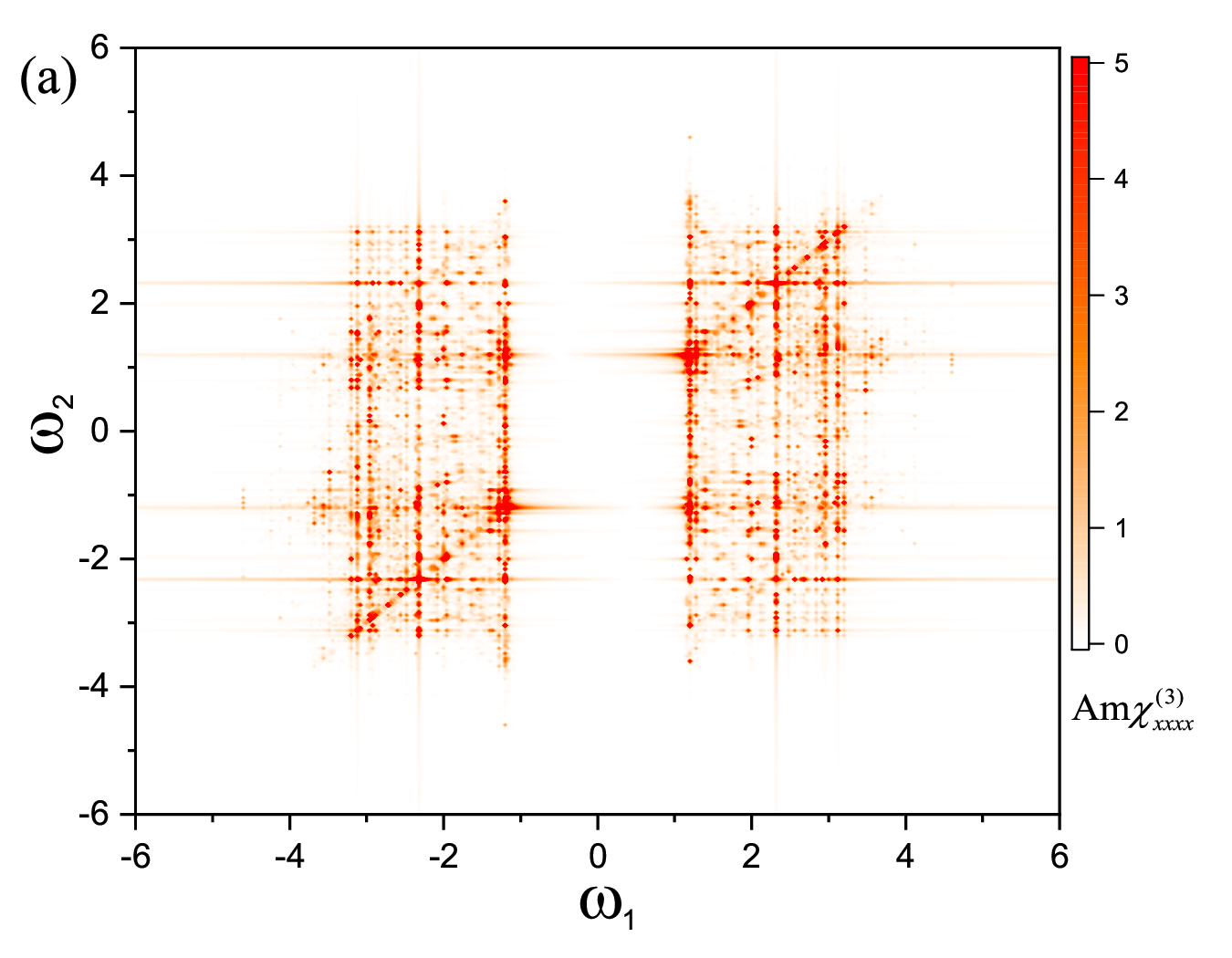}  
 	\includegraphics[angle=0, width=0.45 \columnwidth]{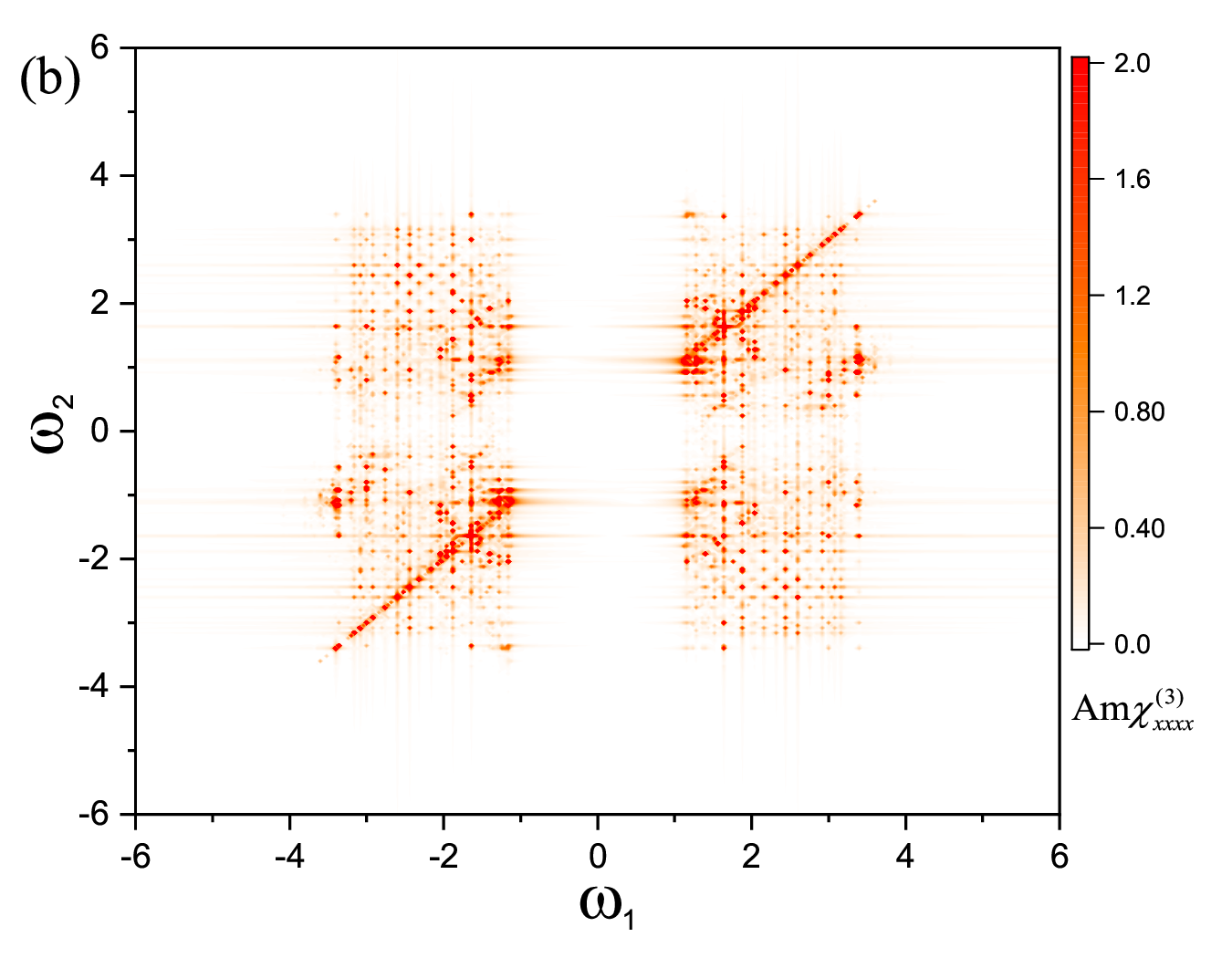}  
 	\caption {(Color online) 
 		Two-dimensional amplitude spectrum of the third-order susceptibilities
 		 $\chi^{(3)}_{xxxx}(\omega_{2},0,\omega_{1})$
 		with chain length $L$=16(a)  and $L$=20(b).}  
 	\label{fig12:tdns_scaling}
 \end{figure}

  For  comparation, we plotted the  the two-dimensional amplitude spectrum of the third-order susceptibilities  
  $\chi_{xxxx}^{(3)}(\omega_{2},0\omega_{1})$ with  chain length $L=16$ and $L=20$.
  The four-point correlation function $C_{R}+iC_{I}$ is calculated based on  
  ORLM. The eigenvalues and eigenstates are obtained by diagonalizing the 
  tridiagonal  Hamiltonian matrix in Krylov space.  
 As shown in Fig. \ref{fig12:tdns_scaling} , the diagonal nonrephase signals are very sharpe.


\end{document}